\documentclass[aps, showpacs, pra, superscriptadaddress, %eqsecnum, twocolumn
] {revtex4-1}
\usepackage{hyperref}
\usepackage{amsmath}
\usepackage{amsfonts} 
\usepackage{amssymb}
\usepackage{epsfig}
\usepackage{natbib}
\usepackage{epstopdf}
\usepackage{graphicx}
\usepackage[normalem]{ulem}

\usepackage{floatrow}
\usepackage[caption=false,font=Large]{subfig}
\usepackage{amssymb}
\usepackage{amsthm}

\newcommand*{\be}{\begin{equation}}
\newcommand*{\ee}{\end{equation}}
\newcommand*{\bea}{\begin{eqnarray}}
\newcommand*{\eea}{\end{eqnarray}}

\begin{document}

\title{Dimer with gain and loss: Integrability and $\mathcal{PT}$-symmetry restoration}

\author{I. V. Barashenkov}
\affiliation{Centre for Theoretical and Mathematical Physics, University of Cape Town, Rondebosch 7701, South Africa and 
Joint Institute for Nuclear Research, Dubna, Russia}

\author{D. E. Pelinovsky}
\affiliation{Department of Mathematics, McMaster University, Hamilton ON, Canada, L8S 4K1 
and Department of Applied Mathematics, Nizhny Novgorod State Technical University, 
Nizhny Novgorod,  Russia
}

\author{P. Dubard}
\affiliation{
Department of Mathematics, University of Cape Town,  Rondebosch 7701, South Africa}

\vspace*{10mm}

%%%% Abstract text to be placed here %%%%%%%%%%%%
\begin{abstract}
A  $\mathcal{PT}$-symmetric nonlinear Schr\"odinger  dimer is  a
two-site discrete nonlinear Schr\"odinger equation
with one site losing and the other one gaining energy at the same rate.
In this paper,  two four-parameter families of cubic  $\mathcal{PT}$-symmetric dimers
are constructed
as  gain-loss extensions of their conservative, Hamiltonian, counterparts.
We prove that   
all these damped-driven 
equations
  define
completely integrable Hamiltonian systems.
The second aim of our study is to identify nonlinearities that  give rise to the spontaneous $\mathcal{PT}$-symmetry restoration.
 When the symmetry of the underlying linear dimer is broken
 and an unstable small perturbation starts to grow,  the
 nonlinear coupling of the required type 
 diverts progressively large amounts of energy from the
gaining to the losing site.
As a result, the exponential growth is saturated and
all trajectories remain  trapped
in a finite part of the phase space regardless of the value of the gain-loss coefficient.

\end{abstract}
%%%%%%%%%%%%%%%%%%%%%%%%%%%


\maketitle

\section{Introduction}

The  nonlinear Schr\"odinger {\it dimer\/} is a code name for the discrete nonlinear Schr\"odinger equation defined on a lattice consisting  just of two sites:
\begin{align}
\begin{split}
i {\dot u} +v    =  F(u,u^*, v,v^*),  \\
i {\dot v}  +u  = G(u,u^*, v,v^*).
\end{split}
\label{D1}
\end{align}
Here and in what follows, the overdot denotes the derivative with respect to $t$.

 The  dimer \eqref{D1} is one of the simplest (and hence most heavily used) models of a coupled bimodal structure.
 In optics, the system \eqref{D1} describes the directional coupler --- a pair of parallel waveguides coupled through their evanescent fields.
In this context, $u$ and $v$ are the  complex amplitudes of stationary light beams in the waveguides and $t$ measures the distance along their parallel cores
 \cite{couplers,Maier,CSP}.

 When the same system is employed
in the studies of the
Bose-Einstein condensates,
 $u$ and $v$ stand for  the  amplitudes of the
  mean-field condensate
wave functions localized in the left and right well  of  a double-well potential \cite{BEC} (or of their symmetric and antisymmetric combinations  \cite{Ostrovskaya}).
In this application, $t$ has the meaning of time.

 The nonlinear Schr\"odinger dimers were also utilised
 in the solid state physics  \cite{Eilbeck,DST,Tsironis_review}
and in the context of electric lattices \cite{Tsironis_review}.

Typically, the dimer arises  as an amplitude
equation; that is, $u$ and $v$ represent slowly changing amplitudes of some oscillatory variables $x$ and $y$:
\[
x(\tau)= u(t) e^{i \omega \tau} + u^*(t) e^{-i \omega \tau} + ... ,
\quad
y(\tau)= v(t) e^{i \omega \tau} +v^*(t) e^{-i \omega \tau} + ... .
\]
Here $t= \epsilon^2 \tau$, $\epsilon^2$ is a small parameter, and the dots stand for small anharmonic corrections. The  invariance of the optical or atomic  structure with respect to translations in $\tau$
is inherited by the  amplitude equations as
the  invariance  under  simultaneous phase shifts in $u$ and $v$.
In other words, physically meaningful nonlinearities have to satisfy
\begin{align} \begin{split} 
F(e^{i \phi}u,  e^{-i \phi} u^*, e^{i \phi}v, e^{-i \phi}v^* )  = & e^{i \phi} F (u,u^*,v, v^*),       \\
G (e^{i \phi}u,  e^{-i \phi} u^*,  e^{i \phi}v,  e^{-i \phi} v^*)  = & e^{i \phi} G (u,u^*, v, v^*)
\label{U1} \end{split} 
 \end{align}
for any real $\phi$.

Another property of the dimer dictated by physics, is  conservation of energy
which gives rise to an underlying Hamiltonian structure.
There are two main types of Hamiltonian formulations admitted by the dimers.
One class of Hamiltonian dimers has the {\it straight-gradient\/} form
\be
i   \frac{du}{dt}=\frac{\partial {\mathcal H}}{\partial {u^*}},\qquad
 i \frac{dv}{dt}= \frac{\partial {\mathcal H}}{\partial {v^*}}.
 \label{D2}
\ee
Here $\mathcal H$ is the Hamilton function while the canonical coordinate-momentum pairs
 are $(u, u^*)$ and $(v, v^*)$.
Alternatively,  $u$ can be paired with $v^*$, while
$u^*$ play the role of
  momentum
conjugate to $v$:
\be
i   \frac{du}{dt}=\frac{\partial {\frak H}}{\partial {v^*}},\qquad
 i \frac{dv}{dt}= \frac{\partial {\frak H}}{\partial {u^*}}.
 \label{D3}
\ee
(This time we use a different notation for the Hamilton function to keep the treatment of the two cases separate.) 
In what follows, we are referring to equations \eqref{D3} as the {\it cross-gradient\/} systems.

The last requirement is that of the left-right symmetry
(the parity symmetry) of the system. This requirement
arises if the two elements making up the dimer
(two waveguides or two potential wells) are identical.
Mathematically, it reduces to the invariance under the permutation of $u$ and $v$ in \eqref{D1}:
\be
F(u, u^*, v, v^*)= G(v, v^*, u, u^*).
\label{F-G-formula}
\ee

If the two elements are not identical
--- for example, if one channel is dissipative while the other one draws  energy from outside ---
 the system
 may be still invariant under  a relaxed form of the left-right permutation
 known as the  parity-time ($\mathcal{PT}$) symmetry \cite{PT}. Mathematically,  the
discrete Schr\"odinger equation is said to be $\mathcal{PT}$-symmetric if it is
invariant under the product of $\mathcal{P}$ and $\mathcal{T}$ transformations.
Here the $\mathcal P$ operator swops the two elements around,
\be
{\mathcal P} \left(
\begin{array}{c}
u  \\ v
\end{array}
\right)
=
 \left(
\begin{array}{c}
v \\ u
\end{array}
\right),
\nonumber
\ee
while the $\mathcal T$  represents the effect of  time inversion on the complex amplitudes:
${\mathcal T} u(t)= u^*(-t)$, ${\mathcal T} v(t)= v^*(-t)$.

The current upsurge of interest in the $\mathcal{PT}$-symmetric systems
is  due to the fact that  they can strike the balance between the gain of
energy in one channel and  loss in the other.
In the optical context, the $\mathcal{PT}$-symmetric dimer describes
a waveguide with power loss coupled to a waveguide experiencing optical gain
\cite{Maier,CSP,PT_Opt,Guo,NatPhys,Ramezani}.
In the matter-wave setting, the $\mathcal{PT}$-symmetric system
is formed by two quantum states, with one state leaking and the other one being fed with particles \cite{Graefe,Heiss}.

Since the $\mathcal{PT}$-symmetric systems have channels for the energy exchange with the
environment, they are commonly thought to occupy a niche between dissipative and
conservative systems. It  was therefore met with surprise when
 some linear \cite{Benfreda}  and nonlinear \cite{BG} $\mathcal{PT}$-symmetric systems were found to
possess Hamiltonian structure.
In particular, there are Hamiltonian $\mathcal{PT}$-symmetric dimers;
one example was produced in \cite{BG}:
\begin{align}
\begin{split}
i {\dot u}     +  v
+  (|u|^2+ 2 |v|^2) u
+  v^2 u^* &   =                      \phantom{-}   i   \gamma u,    % \nonumber
\\
i {\dot v}   +  u
+  (|v|^2+2|u|^2) v
+ u^2 v^*  &  =    - i    \gamma v.
\label{A1}
\end{split}
\end{align}
The two terms in the right-hand sides of \eqref{A1} account for the gain and loss of energy, with
$\gamma>0$  being the   gain-loss coefficient.

Another Hamiltonian $\mathcal{PT}$-symmetric system was identified in \cite{B}:
 \begin{align}
 \begin{split}
i {\dot u}  + v+ |u|^2 u   & =   \phantom{-}   i \gamma u,  \\   %\nonumber  \\
i {\dot v}   +u+ |v|^2v   &  =  -   i\gamma v.
\label{A3}
\end{split}
\end{align}
Because of its ubiquity in physics
\cite{Maier,CSP,NatPhys,Ramezani,Graefe,Turitsyn},
 equation \eqref{A3}  is occasionally referred to as the  {\it standard\/} dimer.
 
Finally, the Hamiltonian model
 \begin{align}
 \begin{split}
i {\dot u}  + v
   - (2 \alpha_1+\alpha_2)|u|^2v - 2 \alpha_1 |v|^2 v
    &   =     \phantom{-} i\gamma u,
     \\   % \nonumber  \\
i {\dot v} +u   -
   (2 \alpha_1 + \alpha_2)|v|^2  u    -   2 \alpha_1 |u|^2 u & =
   -i \gamma v,
\label{J1}
\end{split}
\end{align}
was discovered outside the domain of the $\mathcal{PT}$-symmetry
--- as a by-product in the search 
 of integrable equations \cite{Jorgensen}.
Here $\alpha_1$ and $\alpha_2$ are arbitrary real coefficients.

 The availability of the Hamiltonian structure is a fundamental property of
a dynamical system.   This property by itself implies the conservation of phase volume
and hence some degree of regularity of motion. But in the presence of additional first integrals
it allows to establish an even higher level of regularity, namely,
 the Liouville integrability.

The first aim of this paper is to show that
 {\it any\/} cubic $\mathcal{PT}$-symmetric  phase-invariant dimer
% covariant under the U(1) transformations \eqref{U1},
obtained as
a $\mathcal{PT}$-symmetric extension of the conservative dimer \eqref{D2} or \eqref{D3},
is a Hamiltonian system.  By determining an additional integral of motion independent of the Hamiltonian, we establish
the complete integrability of all these systems.

Another topic pursued in the
 present study concerns the phenomenon  of $\mathcal{PT}$-symmetry breaking
  --- one of the experimentally accessible properties of physical systems with gain and loss \cite{NatPhys,PT2,Moiseyev,PT_breaking}.
  The spontaneous symmetry breaking occurs
  in  {\it linear\/} $\mathcal{PT}$-symmetric systems    as the gain-loss coefficient  is increased
beyond a critical value $\gamma_c$.
   This exceptional point
separates the symmetric phase
   ($\gamma< \gamma_c$), where all perturbation frequencies
are real, and the symmetry-broken phase
   ($\gamma>\gamma_c$), where some
frequencies are complex and the corresponding modes grow
exponentially.

 When the input power in the physical structure is low --- or, equivalently, when the
 initial conditions of the corresponding mathematical model are small --- all nonlinear
 effects are negligible and the system follows the linear laws.
 In particular, small initial conditions
 in the symmetry-broken phase trigger an exponential growth.
 However as the resulting solution reaches finite amplitude,  the nonlinear coupling terms  kick in.
 These terms  may channel the power from the site where it is gained, to the site where it is lost.
% If that mechanism is at work,  the higher is the power gained, the larger portion of this power is channeled to the disposal site.
The higher is the power gained, the larger portion of it is channeled to the disposal site by the nonlinear coupling.
In systems where this mechanism is at work, the exponential growth is arrested and all escaping trajectories are sent back to the finite part of the phase space.
 The $\mathcal{PT}$ symmetry becomes spontaneously restored.

 The classification of  integrable $\mathcal{PT}$-symmetric dimers
 with  the nonlinearly restored $\mathcal{PT}$-symmetry,
  is the second objective of our study.

The outline of the paper is as follows.

In section \ref{cg}  we present a four-parameter family of
the $\mathcal{PT}$-symmetric {\it cross-gradient\/} dimers; all these systems are Hamiltonian in their original $u$ and $v$ variables.
In a similar way, a four-parameter $\mathcal{PT}$-symmetric extension of
 the {\it straight-gradient\/}  dimer is introduced in section \ref{sg}.

The straight-gradient systems do not admit the Hamiltonian formulation in terms of  $u$ and $v$ ---
except when the straight-gradient system is  cross-gradient  at the same time,
or when it is gauge-equivalent to a cross-gradient system. (We identify such dual cases in section \ref{sg}.)
Nevertheless, transforming to the Stokes variables (section \ref{S})
we can describe all trajectories of the straight-gradient dimer and elucidate the geometry of its phase space.

In the subsequent three sections we determine the canonical coordinates
for the general straight-gradient dimer
and reformulate it as a Hamiltonian system.  Three complementary subfamilies of the straight-gradient models
are considered
(sections \ref{om_posi}, \ref{neg} and \ref{sing}).

Section \ref{ST} is concerned with the  trajectory confinement and $\mathcal{PT}$-symmetry restoration.
We identify broad classes of nonlinearities capable of suppressing the
exponential blowup regimes --- both within the cross-gradient and straight-gradient families.

Section \ref{appli} offers examples of simple oscillatory systems with the amplitude equations 
in the form of cross- and straight-gradient dimers.

Finally, in section \ref{conclusions} we summarise mathematical results of this study and discuss their physical implications.

\section{Cross-gradient $\mathcal{PT}$-symmetric  dimer}
\label{cg}

A general  cross-gradient dimer \eqref{D3},
complying with the phase invariance \eqref{U1}, 
with no gain or loss, with cubic nonlinearity,
permutation
property (\ref{F-G-formula})
and linear part of the form \eqref{D1},
 is defined by the
Hamiltonian
\begin{equation}
\label{D4}
\frak{H}_0 = -(|u|^2 + |v|^2)  + W(u,v).
\end{equation}
Here $W$ is a U(1)-invariant real quartic polynomial in $u$, $v$ and their complex-conjugates, 
 which is symmetric with respect to the $u \leftrightarrows v$ permutations.
 The most general quartic  polynomial with these properties can be written as
\begin{equation}
\label{D5}
W = \alpha_1 (|u|^2 + |v|^2)^2 + \alpha_2 |u|^2 |v|^2 +
\alpha_3 (u^* v + u v^*) (|u|^2 + |v|^2) +
\alpha_4 (u^* v + u v^*)^2,
\end{equation}
where $\alpha_1,\alpha_2,\alpha_3$,  and $\alpha_4$ are real coefficients.
(See Section 2 in \cite{chugunova}).

For any set of $\alpha$'s, this gainless lossless system admits a straightforward $\mathcal{PT}$-symmetric extension
\be
i   \frac{du}{dt}=\frac{\partial \frak{H}}{\partial {v^*}},
\qquad
 i \frac{dv}{dt}= \frac{\partial \frak{H}}{\partial {u^*}},
 \label{D6}
\ee
where the Hamilton function $\frak{H}$ is different from  $\frak{H}_0$  in just one term:
\be
\frak{H}   = -(|u|^2 + |v|^2)  + W(u,v) + i \gamma (uv^*-u^*v).
\label{D7}
\ee
Here $\gamma>0$ is the gain-loss coefficient.
Substituting the expression \eqref{D7} with $W$ as in \eqref{D5} in equations \eqref{D6}, we obtain a four-parameter family of
Hamiltonian $\mathcal{PT}$-symmetric cubic dimers:
\begin{align} \begin{split} 
i {\dot u}  + v - i \gamma u & = \alpha_3 (|u|^2+ 2|v|^2) u + \alpha_3 v^2 u^*
+ 2\alpha_4 u^2 v^* + \left[ (2 \alpha_1+\alpha_2+ 2 \alpha_4)|u|^2+ 2 \alpha_1 |v|^2 \right]v,   \\
i {\dot v} +u+ i \gamma v & = \alpha_3(2|u|^2+|v|^2) v + \alpha_3 u^2v^*
+ 2 \alpha_4 v^2 u^* + \left[      (2 \alpha_1 + \alpha_2+ 2 \alpha_4)|v|^2          +   2 \alpha_1 |u|^2 \right]u.
\label{D8}
\end{split}
\end{align}

A particular case of \eqref{D8} is the system \eqref{A1}. This is selected by
 letting $\alpha_3=-1$ and $\alpha_1=\alpha_2=\alpha_4=0$. Another special case is
 the dimer \eqref{J1}; this model corresponds to $\alpha_3=\alpha_4=0$.
 The Hamiltonian structure of these two particular systems has been determined earlier
 \cite{BG,Jorgensen}.

Despite the seeming complexity, all trajectories of \eqref{D8} admit a simple analytic description.
We define the
 Stokes vector $\boldsymbol{\mathcal R}= {\bf i} {X}+  {\bf j} {Y}+  {\bf k} {Z}$, where
 \be
X= u^*v+ v^* u, \quad
Y= i (u^*v-v^*u), \quad
Z= |u|^2-|v|^2.
\label{Stokes}
\ee
Note that the length $\mathcal{R}=\sqrt{X^2+Y^2+Z^2}$ of the Stokes vector has a simple expression in terms of $u$ and $v$:
\[
\mathcal{R} = |u|^2 + |v|^2.
\]

Transforming to $X, Y$,  and $Z$,  equations \eqref{D8} give
a dynamical system in three dimensions:
\begin{subequations} \label{Cr}
\begin{align}
\dot{X} = 0,
 \label{C1}  \\
\dot{Y} =-2Z + 2 \alpha_3 XZ + 4 \alpha_1 Z \mathcal R,
 \label{C2}   \\
\dot{Z} =  2Y+ 2 \gamma \mathcal R - 2 \alpha_3 XY -  (\alpha_2+4 \alpha_1)Y \mathcal R.
\label{C3}
\end{align}
\end{subequations}
%{\color{blue} DROP THE FOLLOWING RED TEXT:}
%{\color{red} We also note a useful consequence of the system \eqref{Cr}:}
%\be
%{\color{red} \dot{\mathcal R} = 2 \gamma Z - \alpha_2 YZ.}
%\label{C4}
%  \ee

Equation \eqref{C1} implies
 that for any selection of $\alpha_1$, $\alpha_2$, $\alpha_3$, and $\alpha_4$,
the cross-gradient dimer  \eqref{D8} has two independent integrals of motion: $X$ and $\frak H$.
Accordingly, the Hamiltonian system \eqref{D8} is Liouville-integrable.

%{\color{blue} DROP THE FOLLOWING RED PARAGRAPH AND REPLACE IT BY THE FOLLOWING BLUE PARAGRAPH.} 
%{\color{red} 
%All trajectories lie in parallel vertical planes $X =X_0=const$.
%To find the form of the trajectories we divide
%the left- and right-hand side of equation \eqref{C2} by the the left- and right-hand side of
%equation \eqref{C4}, respectively. Considering $Y$ as a function of $\mathcal R$ this gives
%a separable equation
%\[
%\frac{dY}{d \mathcal R} = \frac{2 \alpha_3 X_0 + 4 \alpha_1 \mathcal R -2}{2 \gamma - \alpha_2 Y}
% \]
% with an implicit solution
% \be
% \frac{\alpha_2}{2} Y^2 + 2 \alpha_3 X_0 \mathcal R+ 2\alpha_1 {\mathcal R}^2 - 2(\gamma Y + \mathcal R) = C.
% \label{C5}
% \ee
% Here
% $C$ is a constant of integration.
% Equation \eqref{C5} with  $\mathcal R= \sqrt{X_0^2+ Y^2+Z^2}$
% defines a one-parameter family of trajectories 
% on the $X=X_0$ plane.
% }

% {\color{blue}
All trajectories lie in parallel vertical planes $X =X_0$, where $X_0$ is an arbitrary constant.
The form of the trajectories is determined by setting $X=X_0$ in the 
equation $\frak{H}(X,Y,Z)=const$, with $\frak{H}$  being the Hamiltonian \eqref{D7} expressed in the Stokes variables:
\be
Y \left( \frac{\alpha_2}{4} Y-\gamma \right)  +  \mathcal{R} \left( \alpha_1 {\mathcal R}+ \alpha_3 X_0-1 \right)  = C.
\label{C5}
\ee
Here
$C$ is a constant of integration.
Equation \eqref{C5} with  $\mathcal R= \sqrt{X_0^2+ Y^2+Z^2}$
defines a one-parameter family of trajectories 
on the $X=X_0$ plane.
% }

\section{Straight-gradient $\mathcal{PT}$-symmetric dimer}
\label{sg}

The conservative straight-gradient dimer
with general cubic nonlinearity, left-right symmetry and phase invariance,
has the form
\be
i   \frac{du}{dt}  =\frac{\partial {\mathcal H}}{\partial {u^*}},\qquad
 i \frac{dv}{dt} = \frac{\partial {\mathcal H}}{\partial {v^*}},
 \label{D9}
\ee
where
\be
\mathcal H= -(uv^*+u^*v) +W(u,v),
\label{D0}
\ee
and
$W$ is the  four-parameter quartic  polynomial:
\begin{equation}
\label{D500}
W = \beta_1 (|u|^2 + |v|^2)^2 + \beta_2 |u|^2 |v|^2 +
\beta_3 (u^* v + u v^*) (|u|^2 + |v|^2) +
\beta_4 (u^* v + u v^*)^2.
\end{equation}
Here  $\beta_1,\beta_2,\beta_3$,  and $\beta_4$ are real coefficients. 
This is the same quartic as in the previous section; 
we have just switched from the  $\alpha$-
to the $\beta$-notation  to emphasise that we consider a totally new family of models.

The system \eqref{D9}-\eqref{D500}  with $\beta_1=-\frac12$, $\beta_2=1$, and $\beta_3=\beta_4=0$
is known as the $N=2$ discrete self-trapping equation \cite{Eilbeck}.
 When $\beta_3=\beta_4=0$ while
$ \beta_1=-\rho/2$, $\beta_2=\rho-1$ with $\rho$  a real coefficient,
these equations constitute the spatially homogeneous version of the Aceves-Wabnitz coupled mode system
for the  nonlinear optical  grating \cite{AW}.
A particular case of this ($\beta_1=0$,
$\beta_2=-1$) is the spatially-independent   Thirring model (a theory of self-interacting spinor field)
\cite{MTM,BSV,David}.
Another case related to spinors is $ \beta_4=-\frac12$,  $\beta_1 = \beta_2 = \beta_3=0$; 
this system derives from the one-component Gross-Neveu model \cite{GN,BSV}.
The system with $\beta_2=2$, $\beta_4=- \frac12 $, $\beta_1=\beta_3=0$ is related to
 the spinor theory with the pseudoscalar interaction \cite{Lee,David}.

The $\mathcal{PT}$-symmetric extension of the general straight-gradient dimer \eqref{D9}
has the form
\be
i   \frac{du}{dt}  - i \gamma u =\frac{\partial {\mathcal H}}{\partial {u^*}},\qquad
 i \frac{dv}{dt} + i \gamma v= \frac{\partial {\mathcal H}}{\partial {v^*}}.
 \label{D90}
\ee
Evaluating the partial derivatives using \eqref{D0} and \eqref{D500}, these equations become
\begin{align} \begin{split}
i{\dot u} +v-i \gamma u & =
\left[ 2 \beta_1 |u|^2+  (2 \beta_1+ \beta_2 + 2 \beta_4)|v|^2 \right]  u+    2 \beta_4 v^2 u^*  + \beta_3 u^2v^*   +\beta_3 \left(2 |u|^2  + |v|^2 \right) v,  \\
i {\dot v} +u+ i \gamma v & =
\left[ 2 \beta_1 |v|^2 +(2 \beta_1+\beta_2+ 2 \beta_4) |u|^2 \right] v +  2 \beta_4 u^2 v^* + \beta_3v^2u^*
+\beta_3\left(  2 |v|^2      +    |u|^2      \right)     u.
\label{D10}
\end{split}
\end{align}
The {\it standard\/} dimer \eqref{A3} is a special case of \eqref{D10}. This is selected by taking
$\beta_1=-\frac12$, $\beta_2=1$, and $\beta_3=\beta_4=0$ in equations \eqref{D10}.
There is an extensive literature on mathematical aspects of this model
\cite{Ramezani,SXK,SP,SP1,pel1,flach,susanto,pel2,B}.

The couples $(u, u^*)$ and $(v, v^*)$ do not form pairs of canonically conjugate variables.
That is, the straight-gradient dimer \eqref{D10} does not admit a  Hamiltonian formulation in terms of the
original complex coordinates --- except when the straight-gradient dimer is  cross-gradient at the same
time. The necessary and sufficient condition for the equation \eqref{D9} to have a representation \eqref{D6}
with some $\frak{H}$, is
\[
\frac{\partial }{\partial u^*}\left(        i \gamma u + \frac{\partial {\mathcal H}}{\partial u^*} \right)  = \frac{\partial }{\partial v^*}
\left(  - i \gamma v + \frac{\partial {\mathcal H}}{\partial v^*} \right).
\]
For $\mathcal{H}$ of the form \eqref{D0},  this condition translates into
\[
\frac{\partial^2 W}{\partial{ u^*}^2} = \frac{\partial^2 W}{\partial {v^*}^2}.
\]
Substituting the quartic polynomial \eqref{D500} for $W$ gives, finally, $\beta_1=\beta_4$.

The choice $\beta_1=\beta_4$ ensures the existence of a (complex) function $\frak H$ such that
\begin{align}
i \gamma u + \frac{\partial {\mathcal H}} {\partial u^*}  & =  \frac{\partial {\frak H}}{\partial v^*},
\quad
-i \gamma u^* + \frac{\partial {\mathcal H}} {\partial u}    =   \frac{\partial {\frak H}^*}{\partial v},
\label{D31}
 \\
-i \gamma v + \frac{\partial {\mathcal H}} {\partial v^*} & =   \frac{\partial {\frak H}}{\partial u^*},
\quad
i \gamma v^* + \frac{\partial {\mathcal H}} {\partial v}    =  \frac{\partial {\frak H}^*}{\partial u}.
\label{D32}
\end{align}
The necessary and sufficient condition for the function $\frak H$  in \eqref{D31} to be real, is given by
\[
\frac{\partial}{\partial v} \left( i \gamma u+ \frac{\partial \mathcal{H}}{\partial u^*}  \right)=
\frac{\partial }{\partial v^*} \left( - i \gamma u^* + \frac{\partial \mathcal{H}} {\partial u} \right).
\]
In a similar way, the necessary and sufficient condition for ${\frak H} ={\frak H}^*$ in
equation \eqref{D32} is
\[
\frac{\partial}{\partial u} \left(- i \gamma v + \frac{\partial \mathcal{H}}{\partial v^*}  \right)=
\frac{\partial }{\partial u^*} \left(  i \gamma v^* + \frac{\partial \mathcal{H}} {\partial v} \right).
\]
For $\mathcal{H}$ of the form \eqref{D0}, each of these two conditions amounts to
\[
\frac{\partial^2 W}{\partial     v^*  \partial
u }= \frac{\partial^2 W}{\partial u^* \partial v},
\]
which gives $2 \beta_1+ \beta_2= 2 \beta_4$.
Using the previously established condition $\beta_1=\beta_4$, this relation reduces simply to $\beta_2=0$.

Thus, the straight-gradient dimer \eqref{D10} with $\beta_1=\beta_4$ and $\beta_2=0$  is,
at the same time,  a Hamiltonian system
with the cross-gradient canonical structure \eqref{D6}.
The  corresponding Hamilton function
$\frak H$ is determined by simple integration:
\be
%\begin{split}
\frak H=
-(|u|^2 + |v|^2)  +
 \frac{\beta_3}{2} (|u|^2+|v|^2)^2 + 2 \beta_4(|u|^2+|v|^2) (uv^*+u^*v)   % \\
  + \frac{\beta_3}{2} (uv^*+ vu^*)^2+
i  \gamma (uv^*-vu^*).
\label{H-cr}
%\end{split}
\ee
The Hamiltonian \eqref{H-cr} is of the form  \eqref{D7},  \eqref{D5}
with $\alpha_1=\alpha_4=\frac12 \beta_3$, $\alpha_2=0$, and $\alpha_3=2 \beta_4$.
%[The tilde over ${ H_{\rm cg}}$ serves to emphasise that the  Hamiltonian \eqref{H-cr} {\it does not\/} coincide with the one in \eqref{D5}, \eqref{D7}.]

In fact, the class of straight-gradient dimers admitting the Hamiltonian formulation is even wider.
Let $u_0$ and $v_0$ be a solution to the straight-gradient equations (\ref{D10})
with $\beta_2=0$ and {\it generic\/} $\beta_1, \beta_4$
(that is, $\beta_1$ not necessarily coinciding with $\beta_4$). The
gauge transformation
\begin{align}
u_0= e^{i \varphi} u, \quad
v_0= e^{i \varphi} v \nonumber
\label{D101}
\end{align}
 with 
\[ \varphi(t) = 2 (\beta_4-\beta_1)  \int_0^t (|u|^2 + |v|^2) d \tau \]
generates functions $u$ and $v$ which satisfy the straight-gradient dimer equations with $\beta_2=0$
and $\beta_1$ set equal to $\beta_4$:
\begin{align}
\begin{split}
i{\dot u} +v-i \gamma u & =2 \beta_4
\left( |u|^2+  2 |v|^2 \right)  u+    2 \beta_4 v^2 u^*  + \beta_3 u^2v^*   +\beta_3 \left(2 |u|^2  + |v|^2 \right) v,  \\
i {\dot v} +u+ i \gamma v & =2 \beta_4
\left(  |v|^2 + 2 |u|^2 \right)  v +  2 \beta_4 u^2 v^* + \beta_3v^2u^*
+\beta_3\left(  2 |v|^2      +    |u|^2      \right)     u.
\end{split}
\label{D100}
\end{align}
As we already know, this system has a cross-gradient Hamiltonian formulation with the
Hamilton function \eqref{H-cr}.
Therefore,  the straight-gradient dimer with $\beta_2 = 0$ {\it and any value\/}  of   $(\beta_4-\beta_1)$
is gauge-equivalent to the cross-gradient Hamiltonian system
\eqref{D6},  where the Hamilton function $\frak H$ is as in (\ref{H-cr}).

 In what follows, we uncover the Hamiltonian formulation of  the straight-gradient system 
(\ref{D10})  with $\beta_2 \neq 0$.
This class will include, in particular, the standard dimer \eqref{A3} (which has $\beta_2=1$).

\section{Phase space of  straight-gradient dimer}
\label{S}

The generic ($\beta_2 \neq 0$) straight-gradient dimer does not admit the Hamiltonian formulation
in terms of the original $u$ and $v$ variables.
In order to determine the canonical pairs of coordinates, we transform it to the
 Stokes variables  \eqref{Stokes}.
Equations \eqref{D10} give
a dynamical system in three dimensions:
\begin{subequations} \label{St-1}
\begin{align}
\dot{X} = - \beta_2 YZ,   \label{S1}  \\
\dot{Y} = - 2 Z + (\beta_2 + 4 \beta_4) XZ + 2 \beta_3 Z \mathcal{R},     \label{S2}   \\
\dot{Z} = 2 Y + 2\gamma \mathcal{R} - 4 \beta_4 XY  - 2 \beta_3 \mathcal{R} Y.
\label{S3}
\end{align}
\end{subequations}
We also note an equation for the length of the Stokes vector that follows from
 the system \eqref{St-1}:
\be
\dot{\mathcal R} = 2  \gamma Z.
\label{S4}
\ee

Replacing $t$ with  $\mathcal R$ as a new independent variable, and using 
$d/dt= {\dot {\mathcal R}} \, d/d \mathcal R$,  equations \eqref{S1} and \eqref{S2} become a linear nonhomogeneous system
% for $X(\mathcal{R})$ and $Y(\mathcal{R})$:
\begin{align} \begin{split}
2 \gamma \frac{dX}{d \mathcal{R}} + \beta_2 Y=0,  \\
2 \gamma \frac{dY}{d \mathcal{R}}  -  (\beta_2+ 4 \beta_4) X  = -2 + 2 \beta_3 \mathcal{R}.
\label{D11}
\end{split}
\end{align}
Our strategy will be to determine the general solution of \eqref{D11}
including two constants of integration. These ``constants of motion" of the system \eqref{D11}
will serve as the two first integrals of the original three-dimensional dynamical system \eqref{St-1}.
To obtain the Hamiltonian formulation of the dimer \eqref{D10},
one of these will be appointed as the Hamiltonian and the other one as a canonical coordinate.

As in the case of the cross-gradient dimer,
the existence of two independent conserved quantities
along with the availability of the Hamiltonian structure will imply that 
 the straight-gradient dimer \eqref{D10} is Liouville integrable.

It is convenient to introduce the quantity
\be
\omega^2 \equiv \beta_2(\beta_2+ 4 \beta_4).
\ee
The form of the solution of the system \eqref{D11} depends on whether $\omega^2$ is positive, negative or zero.
We
consider these three cases separately.

%\section{Geometry of the straight-gradient phase space:  $\omega^2>0$}
%\label{geo_straight}

Assuming that $\omega^2 >0$,
the general solution of \eqref{D11} is
\begin{subequations} 
\begin{align}
X= A  \cos \left( \frac{\omega}{2 \gamma} \mathcal{R}  \right)   + B \sin \left( \frac{\omega}{2 \gamma} \mathcal{R}  \right)
+ \frac{ 2- 2 \beta_3 \mathcal{R}}{\beta_2+ 4 \beta_4},
\label{DX}  \\
Y= \frac{\omega}{ \beta_2}
\left[
 A \sin   \left( \frac{\omega}{2 \gamma} \mathcal{R}  \right)  -B     \cos  \left( \frac{\omega}{2 \gamma} \mathcal{R}  \right)   \right]
+ \frac{4 \gamma \beta_3}{\omega^2},
\label{DX2}
\end{align}
   where $A$ and $B$ are constants of integration.
Treating $\mathcal R$  as a parameter, and supplementing \eqref{DX}-\eqref{DX2}
with the formula
\be
Z= \pm \sqrt{{\mathcal R}^2- X^2(\mathcal R)  - Y^2(\mathcal R)},
\label{DX3}
\ee
\label{D12}
\end{subequations}
these equations provide explicit parametric expressions for  trajectories of the dimer: 
$X=X(\mathcal R)$, $Y=Y(\mathcal R)$, $Z=Z(\mathcal R)$.

\begin{figure}
\begin{center}
\includegraphics[height=61mm, width=0.42\linewidth]{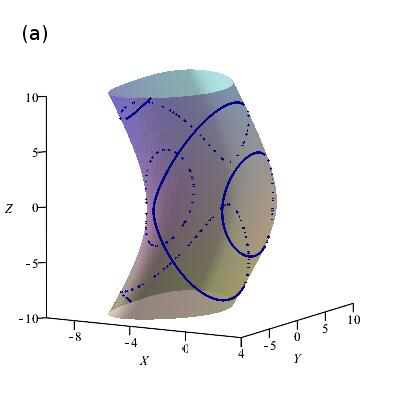}
\includegraphics[height=61mm, width=0.42\linewidth]{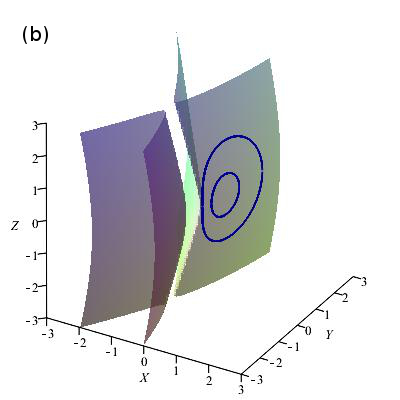}
\includegraphics[height=61mm, width=0.42\linewidth]{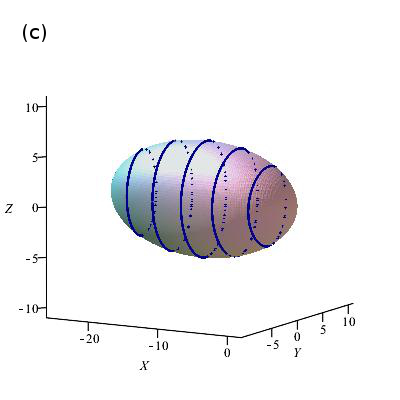}
\includegraphics[height=61mm, width=0.42\linewidth]{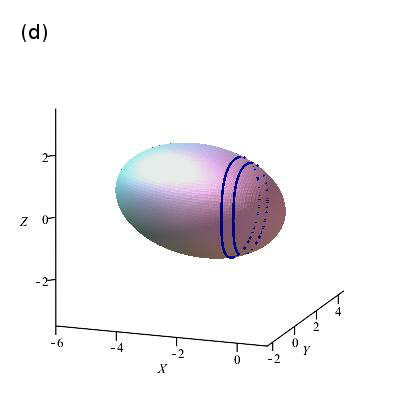}
\includegraphics[height=61mm, width=0.42\linewidth]{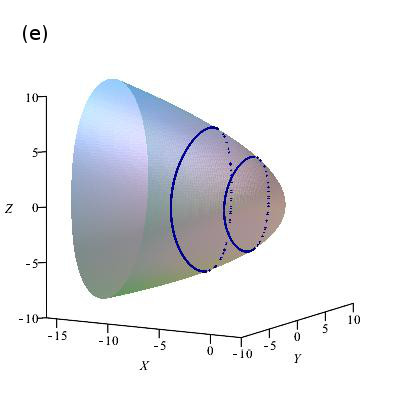}
\includegraphics[height=61mm, width=0.42\linewidth]{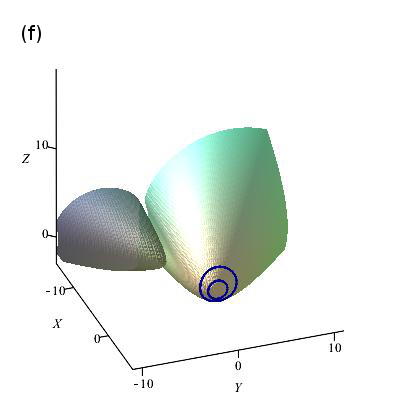}
\end{center}
\caption{\label{F1} 
The surface \eqref{rho} with $\omega^2>0$  (a, c, e) and $\omega^2<0$  (b, d, f).
 The parameter $|\sigma|$ is smaller than 1 in (a,b);
  greater than 1 in (c,d), and equal to 1 in  (e,f).
%  The value of $\beta_2$ is $1$ in (a,b,c) and $-1$ in (d,e,f).
  The value of $\beta_2$ is 1 in (a,c,e)
and -1 in (b,d,f).
The value of $\beta_3$ is 2 in (a),  1 in (b),
5 in (c),
9/2 in (d,e), and 7/2 in (f).
 In all panels,  $\gamma=1$  and $\beta_4=2$.
 The surface parameter $\rho^2=7$ in (a,c,e) and
$s_1 s_2 \rho^2= -\frac{1}{100}$  in (b,d,f).
We also show a few trajectories of the system \eqref{St-1}
lying on each surface.
}
\end{figure}

Denoting $\rho= \sqrt{A^2+B^2}$,
equations \eqref{DX} and \eqref{DX2}  give
\be
\rho^2=
 \left( X+ \frac{2 \beta_3 \mathcal{R} -2}{\beta_2+ 4 \beta_4}  \right)^2
+
\left(  \frac{\beta_2}{\omega} \right)^2  \left( Y-\frac{4 \gamma \beta_3}{\omega^2} \right)^2.
\label{rho}
\ee
 Equation \eqref{rho} with 
 $\mathcal R =   \sqrt{X^2 + Y^2 + Z^2}$  is an implicit equation of the surface on which all trajectories lie.
The shape of this surface in the  $(X,Y,Z)$  
phase space depends on the value of the parameter
\be
\sigma = \frac{ 2 \beta_3}{\beta_2+ 4 \beta_4}.
\label{sigma}
\ee
 Figs.\ref{F1} (a), (b),  and (c)  depict the surface with $|\sigma| < 1$, $|\sigma| > 1$, and $|\sigma| = 1$.

In the case where $ \omega^2  <0$, we define
$\nu^2= -\omega^2>0$. The
 general solution of \eqref{D11} is then
  \begin{subequations}
\begin{align}
X= C  \exp \left( \frac{\nu}{2 \gamma} \mathcal{R}  \right)   + D \exp   \left(- \frac{\nu}{2 \gamma} \mathcal{R}  \right) 
+ \frac{ 2- 2 \beta_3 \mathcal{R}}{\beta_2+ 4 \beta_4}, \label{D91} \\
Y= -\frac{\nu}{ \beta_2}
\left[
 C \exp  \left( \frac{\nu}{2 \gamma} \mathcal{R}  \right)  -D   \exp  \left(- \frac{\nu}{2 \gamma} \mathcal{R}  \right)   \right]
- \frac{4 \gamma \beta_3}{\nu^2},
\label{D920}
\end{align}
\label{D92}
\end{subequations}
% \begin{subequations}
%\begin{align}
%X= A  \cosh \left( \frac{\nu}{2 \gamma} \mathcal{R}  \right)   + B \sinh \left( \frac{\nu}{2 \gamma} \mathcal{R}  \right)
%+ \frac{ 2- 2 \beta_3 \mathcal{R}}{\beta_2+ 4 \beta_4}, \label{D91} \\
%Y= -\frac{\nu}{ \beta_2}
%\left[
% A \sinh   \left( \frac{\nu}{2 \gamma} \mathcal{R}  \right)  +B     \cosh  \left( \frac{\nu}{2 \gamma} \mathcal{R}  \right)   \right]
%- \frac{4 \gamma \beta_3}{\nu^2},
%\label{D920}
%\end{align}
%\label{D92}
%\end{subequations}
where $C$ and $D$ are two constants of integration.

Expressing  $4CD$ in terms of $X,Y, \mathcal{R}$
and denoting it $s_1 s_2   \rho^2$, where $s_1= \mathrm{sign} \, C$ 
and $s_2 = \mathrm{sign} \, D$, we  obtain
\be
s_1  s_2 \rho^2 =
\left( X+ \sigma \mathcal{R} -\frac{2}{\beta_2+ 4 \beta_4}  \right)^2
-  \left( \frac{\beta_2}{\nu} \right)^2 \left( Y+\frac{4 \gamma \beta_3}{\nu^2} \right)^2.
\label{D61}
\ee
Here $\rho \geq 0$ and $s_1  s_2$
is either $1$ or $-1$, depending on whether $CD>0$ or $CD<0$.
The solution surface \eqref{D61} made up by all trajectories of the dimer, is
shown in Fig \ref{F1} (d), (e),  and (f)  for $|\sigma| < 1$, $|\sigma| > 1$, and $|\sigma| = 1$, respectively. 

Finally, the case $\omega^2=0$ is considered in section \ref{sing} below.

\section{Hamiltonian structure of straight-gradient dimer: $\omega^2>0$}
\label{om_posi}

It is convenient to write  the explicit solution    \eqref{DX}-\eqref{DX2}  in the complex form
\be
X+i \frac{\beta_2}{\omega} Y= \rho e^{-i \theta} + \frac{ 2- 2 \beta_3 \mathcal{R}}{\beta_2+ 4 \beta_4}
+ i \frac{4 \gamma  \beta_2 \beta_3}{\omega^3},
\label{cmpl}
\ee
 where
\be
\rho e^{-i \theta}= (A-iB) \exp \left( i \frac{\omega}{2 \gamma} {\mathcal R} \right)
\label{YZ1} 
\ee
and $\rho=\sqrt{A^2+B^2}$ was introduced in the previous section.
Assume  $\rho>0$; in this case, equation \eqref{YZ1} serves as the definition of $\theta$.
(We will examine the possibility $\rho=0$ in subsection \ref{rho0}  below.)
 Using equation \eqref{cmpl}, $\theta$ is expressible as a function of $X$, $Y$, and $\mathcal R$:
\be
 \theta = \arctan  \left(  \frac{\omega}{\beta_2}
\frac{X+ \sigma \mathcal R - 2 \beta_2 \omega^{-2} }
{Y-  4 \gamma \beta_3  \omega^{-2}}    \right).
\label{thett}
\ee

The definition \eqref{YZ1} implies that the complex quantity 
\[
\rho   \exp  \left\{  \frac{i}{2 \gamma}  \left( - 2 \gamma \theta - \omega {\mathcal R}  \right) \right\}=A-iB
\]
is a constant.
Therefore its argument 
% Defining an angular coordinate $\theta$ such that
% \begin{subequations}
%\begin{align}
%\begin{split}
% A  \cos \left( \frac{\omega}{2 \gamma} \mathcal{R}  \right)   + B \sin \left( \frac{\omega}{2 \gamma} \mathcal{R}  \right) = \rho \sin  \theta,  
% \label{ABt1} \\
 % A \sin   \left( \frac{\omega}{2 \gamma} \mathcal{R}  \right)  -B     \cos  \left( \frac{\omega}{2 \gamma} \mathcal{R}  \right)= \rho \cos \theta, 
%  \label{ABt2}
 %\end{split}
% \end{align}
% \label{ABt}
%\end{subequations}
 %we can express the constants $A$ and $B$ as
% \[
 %A= \rho \sin \left( \frac{   2 \gamma \theta+ \omega \mathcal R}{2 \gamma}  \right),
% \quad
% B= -\rho \cos \left( \frac{2 \gamma \theta+ \omega \mathcal R}{2 \gamma} \right).
% \]
%Therefore
\be
H=-2\gamma\theta-\omega\mathcal{R}
\label{D26}
\ee
 is a first integral of the system (\ref{St-1}). We shall demonstrate that $H$ can serve as a Hamilton function for the
three-dimensional system \eqref{St-1}
and the underlying nonlinear Schr\"odinger dimer \eqref{D10}.

One more integral of motion,
 $\rho$, is  given by (\ref{rho}).
Choosing $\rho$ and  $\theta$ as a pair of coordinates in the phase space,
we  require that  $\mathcal R$ (and hence $H$)
be a function of $\rho$, $\theta$, and $P_\theta$ --- but not depend on $P_\rho$ \cite{B}.
Irrespectively of how we define $P_\rho$,  the Hamilton equation
\be
{\dot \rho} = \frac{\partial H}{\partial P_\rho} \label{dro}
\ee
will  then reproduce the equation for $\rho$:  ${\dot \rho}=0$.
Another consequence of requiring $\partial H/ \partial P_\rho=0$, is that 
the variable
$P_\rho$
 will not participate in the dynamics  and
the Hamilton equation
\be
{\dot P_\rho } = - \frac{\partial H}{\partial \rho}
\label{Prh}
\ee
will decouple from the rest of the system.

 % The angular  variable  $P_\rho$  will be obtainable from \eqref{Prh} by simple integration.

At this point, we note that, since  $\theta$ is only different from
$-\frac{\omega}{2\gamma} \mathcal{R}$ by a constant,  equation \eqref{S4} gives
\be
{\dot \theta} = - \omega Z.
\label{D20}
\ee
We should choose the momentum $P_\theta$  (denoted ${\mathcal P}$ for brevity) in such a way 
 that
the canonical equation
\be
{\dot \theta} = \frac{\partial H}{\partial {\mathcal P}}
\label{the}
\ee
reproduces equation \eqref{D20}.
For the Hamilton function of the form  \eqref{D26},    equations
\eqref{the} and \eqref{D20} are equivalent if
 \be
 \frac{\partial \mathcal{R}}{\partial {\mathcal P}}=Z.
 \label{D27}
 \ee

Once the equation \eqref{the} is satisfied, the conjugate equation
\be
{\dot {\mathcal P}}= - \frac{\partial H}{\partial \theta}
\label{conj}
\ee
will be verified automatically.
Indeed, since $H$ does not depend on $P_\rho$ while
${\dot \rho}=0$, there are only two nonvanishing terms in 
 the derivative ${\dot H}$:
\[
{\dot H}= \frac{\partial H}{\partial \theta} {\dot \theta} +  \frac{\partial H}{\partial {\mathcal P}} {\dot {\mathcal P}}.
\]
Substituting for $\dot \theta$ from
\eqref{the}, the relation ${\dot H}=0$ implies Eq.\eqref{conj}.

Thus all we need to do in order to put the canonical structure in place,
is to identify
the canonical momentum ${\mathcal P}=P_\theta$ ensuring the validity of equation \eqref{D27}.
Before proceeding to the unveiling of $P_\theta$,
a technical remark is in order.
Equation \eqref{cmpl}  implies
that the coordinate $Y$ can be expressed solely in terms of
$\rho$ and $\theta$ (rather than $\rho$, $\theta$, and $\mathcal{R}$):
\be
Y= \frac{4 \gamma \beta_3}{\omega^2}+ \frac{\omega}{\beta_2} \rho \cos \theta.
\label{YY}
\ee
Therefore $\theta$ may be thought of as the spherical polar angle in the
frame of reference where $Y$ is the vertical coordinate.
In what follows, we also introduce (an analogue of) the azimuthal angle on the $(X,Z)$-plane. This construction will be based on the following
decomposition of the coordinate $X$ stemming from  \eqref{cmpl}:
\be
X= x - \sigma  \mathcal{R}.
\label{D15}
\ee
Here we have isolated a term that is expressible entirely in terms of $\rho$ and $\theta$:
\be
x=\frac{2 }{\beta_2+ 4 \beta_4}  +
 \rho \sin \theta.
\label{X0}
\ee

Our construction of the canonical
momentum $P_\theta$   depends on whether the parameter $\sigma$ defined in
\eqref{sigma} is smaller, greater,   or equal to 1 in magnitude.

\subsection{Hyperbolic case:  $|\sigma|<1$}
\label{hyp}

 Assuming $|\sigma| <1$, we define 
\[
 \Lambda^2 =
1 -\sigma^2 >0.
\]
The corresponding subfamily of models includes, in particular, the standard dimer \eqref{A3}
--- for which $\omega^2=1$ and $\sigma=0$.  (We note that the Hamiltonian structure 
of the standard dimer was elucidated earlier \cite{B}.)

Inserting
 the decomposition \eqref{D15}  in the identity $X^2+Y^2+Z^2= \mathcal{R}^2$ yields
\be
\left( \Lambda \mathcal{R} + \frac{\sigma} { \Lambda} x \right)^2-\left( \frac{\partial {\mathcal R}}{\partial {\mathcal P}} \right)^2
=r^2,
\label{D17}
\ee
where we have substituted for $Z$ from \eqref{D27} and defined
\be
r^2 = \frac{x^2}{\Lambda^2} + Y^2.
\label{Dr}
\ee
Since $x$  and 
$r$ 
are expressible in terms of $\rho$ and $\theta$  only
(i.e., are independent of ${\mathcal P}$),
equation  \eqref{D17} is an ordinary differential equation for ${\mathcal R} ({\mathcal P})$.
 %Therefore, the two terms in the left-hand side of  \eqref{D17} can be parametrised as follows:
%\be
%\label{psi}
%\Lambda \mathcal{R} + \frac{\sigma} {\Lambda} x
%= r \cosh f({\mathcal P}),
%\quad
%Z= r \sinh f({\mathcal P}).  % \label{E22}
%\ee
%The function $f({\mathcal P})$  in \eqref{psi}  should be chosen so that the equation \eqref{D27} is satisfied.
%{\color{red} This derivation needs to be shortened. Replace $Z$ with $\partial \mathcal{R} /\partial {\mathcal P}$ in \eqref{D17};
%then arrive at \eqref{R10} without having to introduce the function $f$.}
%The expression for $\mathcal{R}$ is obtained from the first equation
% in \eqref{psi}:
% \be
% \mathcal{R}= \frac{r}{\Lambda}\cosh f({\mathcal P}) - \frac{\sigma}{ \Lambda^2} x.
% \label{DR}
% \ee
 %Since
% $r$ and $x$  are independent of ${\mathcal P}$,
% the differentiation of \eqref{DR} with respect to this variable is trivial and \eqref{D27} reduces to
%\[
%\frac{df}{d {\mathcal P}} = \Lambda.
%\]
%Hence $f({\mathcal P})$ can be chosen to be simply $\Lambda {\mathcal P}$.
%Thus we have introduced the momentum $P_\theta={\mathcal P}$ such that
To define ${\mathcal P}$, it is sufficient to pick one solution of this separable equation. We choose
\be
\mathcal{R}({\mathcal P}) = \frac{r}{\Lambda}   \cosh(\Lambda {\mathcal P}) - \frac{\sigma} {\Lambda^2} x;
% \label{R1}
\label{R10}
\ee
the formula \eqref{D27} gives then 
\be
Z= r \sinh( \Lambda {\mathcal P}),
\label{R110}
\ee
so that 
\be
P_\theta={\mathcal P}= \frac{1}{\Lambda} \mathrm{arcsinh} \left( \frac{Z}{r} \right).
\label{R100}
\ee

Together with the definition of the coordinate $\rho$ in \eqref{rho}, the coordinate  $\theta$
in \eqref{thett},
the definition \eqref{R100}  completes the set of three canonical variables.
The fourth coordinate, $P_\rho$, can be reconstructed from equation \eqref{Prh}
by  the integration of its right-hand side in $t$.

Summarising, we have cast the straight-gradient dimer \eqref{D10} with $\omega^2>0$ and $\sigma^2 <1$ in the form of a Hamiltonian system
\be
{\dot \rho}= \frac{\partial H}{\partial P_\rho}, \quad
{\dot P_\rho}=-  \frac{\partial H}{\partial \rho}, \quad
{\dot \theta}= \frac{\partial H}{\partial P_\theta}, \quad
{\dot P_\theta}=-  \frac{\partial H}{\partial \theta},
\label{Hams} \ee
with the Hamilton function
\[
H(\rho, \theta, P_\theta)= -2 \gamma \theta - \frac{\omega}{\Lambda^2} 
\left[
\Lambda r  \cosh ( \Lambda P_\theta ) - \sigma x
 \right].
 \]
 Here $x=x(\rho, \theta)$ and $r=r(\rho, \theta)$ are as in \eqref{X0} and \eqref{Dr}, respectively; the 
 variable $Y$ in \eqref{Dr} is defined in \eqref{YY}.

\subsection{Elliptic case: $|\sigma|>1$}
\label{trig}

Assuming that $\sigma^2 >1$ and defining
%\be
\[
 \Omega^2 =
\sigma^2-1>0,
\]
%\label{D40}
%\ee
 the identity $X^2+Y^2+Z^2=\mathcal R^2$ becomes an ordinary differential equation for $\mathcal{R}({\mathcal P})$:
\be
\left( \Omega \mathcal{R} - \frac{\sigma} { \Omega} x \right)^2+\left( \frac{\partial \mathcal{R}}{\partial {\mathcal P}} \right)^2=r^2. \label{D18}
\ee
Here  we used \eqref{D27},
while
the variable $r$ was introduced differently from \eqref{Dr}:
% \be
\[
r^2=
\frac{x^2}{\Omega^2} - Y^2.
\]
%\label{Drr}
%\ee

The separable equation \eqref{D18} is solved
by taking
 \be   \label{R20}
\mathcal{R}= -\frac{r}{\Omega}\cos(\Omega {\mathcal P}) + \frac{\sigma}{ \Omega^2} x;  % \label{R2}
\ee
hence
\be
Z= r \sin( \Omega {\mathcal P}).
\label{R120}
\ee
This provides a simple expression for  the canonical momentum:
 $P_\theta={\mathcal P}= \Omega^{-1} \arcsin (Z/r)$.

Thus, the  straight-gradient dimer \eqref{D10} with $\omega^2>0$ and $\sigma^2 >1$
is cast in the form \eqref{Hams} with 
\[
H(\rho, \theta, P_\theta) =-2 \gamma \theta + \frac{\omega}{\Omega^2} 
\left[ \Omega r  \cos ( \Omega P_\theta) -\sigma x 
\right],
\]
where $x=x(\rho, \theta)$ and $r=r(\rho, \theta)$.

\subsection{Parabolic case: $|\sigma|=1$}

Assuming $\sigma = \pm 1$,
the identity $X^2+Y^2+Z^2=\mathcal{R}^2$ reduces to an equation
\be
x^2 + Y^2 + \left( \frac{\partial \mathcal{R}}{\partial {\mathcal P}} \right)^2= 2 \sigma x  \mathcal{R},
\label{D19}
\ee
with a solution
%\begin{subequations}
\be
\mathcal{R}=\frac{\sigma}{2}   x\left({\mathcal P}^2+1+\frac{Y^2}{x^2}\right).
\label{R3}
\ee
The rule \eqref{D27} gives then $Z=\sigma x {\mathcal P}$, so that 
\be
  {\mathcal P}= \sigma \frac{Z}{x}.
 \label{R30} 
\ee

The construction of the canonical variables in the $\sigma = \pm 1$ sector is hereby complete.
In the canonical equations \eqref{Hams}, the Hamiltonian is
\[
H(\rho, \theta, P_\theta) = -2 \gamma \theta - \frac{ \omega \sigma x}{2} 
\left[
P_\theta^2+1 + \frac{Y^2}{x^2} \right],
\]
with $x=x(\rho, \theta)$ and $Y=Y(\rho, \theta)$.

\subsection{One-dimensional motion: $\rho=0$}
\label{rho0}

Finally, we consider the situation where $A=B=0$ in \eqref{DX}-\eqref{DX2}
and hence, 
 $\rho=0$. Equation \eqref{rho} gives 
 \begin{align}
 X= \frac{2}{\beta_2+ 4 \beta_4}- \sigma \mathcal R,   \label{Q02} \\
 Y= \frac{4 \gamma \beta_3}{\omega^2}. \label{Q01}
 \end{align}
The invariant manifold defined by  $\rho=0$, consists of a single curve lying in the plane \eqref{Q01}.
The parametric equations for this quadratic curve, with $\mathcal R$ as the parameter, 
are given by \eqref{Q02}, \eqref{Q01} and \eqref{DX3}.

 Using the identity $X^2+Y^2+Z^2={\mathcal R}^2$,
the conserved quantity $-2 \gamma^2Y^2$ can be represented as a function of $X$, $Z$ and $\mathcal R$:
\be
-2 \gamma^2Y^2= 2\gamma^2(X^2+Z^2)-  2\gamma^2{\mathcal R}^2.  \label{Q03}
\ee
Here $X$ is expressible in terms of $\mathcal R$ using \eqref{Q02} and the coefficient $- 2\gamma^2$ was introduced for the later convenience.

We choose \eqref{Q03} as Hamilton's function and
 $\mathcal R$ as the canonical coordinate.  The only term in \eqref{Q03} that remains independent of 
$\mathcal R$, is $2 \gamma^2 Z^2$.
Defining the momentum by $\mathcal P=2 \gamma Z$, 
the Hamilton's function \eqref{Q03} becomes
\be
H= \frac{{\mathcal P}^2}{2}  +U(\mathcal R), 
\quad
U= 2\gamma^2  \left( \frac{2}{\beta_2+4 \beta_4} - \sigma  \mathcal{R} \right)^2 - 2 \gamma^2 {\mathcal R}^2,
\label{H1}
\ee
while the canonical equation ${\dot {\mathcal R}}= \partial H/ \partial {\mathcal P}$ reproduces equation \eqref{S4}.
The conjugate equation ${\dot {\mathcal P}}=- \partial H/ \partial \mathcal R$
is then satisfied automatically, because of $\dot H=0$.

 Thus, the motion along the quadratic curve $\rho=0$ is governed by a Hamiltonian system with one degree of
freedom.

\section{Hamiltonian structure of straight-gradient dimer: $\omega^2<0$}
\label{neg}

\subsection{$\rho>0$:
 two-dimensional motion}

Turning to the dimers with $\omega^2<0$, we continue to employ the first integral $\rho$ as one of the two
canonical variables. This time, $\rho^2$ is defined by \eqref{D61} and equals $4|CD|$.
%We consider the case $\rho>0$ first.
Using   the explicit solution (\ref{D92}),  one can form  linear combinations
\begin{align} \begin{split}
X+ \frac{\beta_2}{\nu} Y  & = s_2 \rho e^{-\theta}  +  \frac{ 2-2 \beta_3 \mathcal{R}}{\beta_2+ 4 \beta_4} 
- \frac{4 \gamma \beta_2 \beta_3}{\nu^3},   \\
X- \frac{\beta_2}{\nu} Y &  = s_1 \rho e^{\theta}   +   \frac{ 2-2 \beta_3 \mathcal{R}}{\beta_2+ 4 \beta_4} 
+ \frac{4 \gamma \beta_2 \beta_3}{\nu^3},
\label{W4}
\end{split}
\end{align}
where we have introduced 
\be
s_1 \rho e^\theta=
 2C\exp \left( \frac{\nu}{2\gamma} \mathcal{R} \right) ,  \quad
s_2 \rho e^{-\theta}=
2D \exp \left(- \frac{\nu}{2\gamma} \mathcal{R} \right).
\label{W1}
\ee
Provided $\rho>0$,
either of these two equations defines 
a real   $\theta$ which we adopt
 as 
the second canonical variable.

Writing equations \eqref{W1} in the form 
\be
s_1 \frac{2C}{\rho} =   \exp
\left(\frac{       2\gamma \theta  -  \nu \mathcal{R}}{2 \gamma}  \right),
\quad
s_2 \frac{2D}{\rho}=   \exp
\left(  \frac{\nu \mathcal{R}                    -2\gamma \theta }{2\gamma}  \right),
\label{W2}
\ee
%introduce $\theta$ such that 
%\begin{align}
%\begin{split}
%A  \cosh \left( \frac{\nu}{2 \gamma} \mathcal{R}  \right)  + B   \sinh \left( \frac{\nu}{2 \gamma} \mathcal{R}  \right)=
%\rho
%\left( \frac{1+\epsilon}{2} \cosh \theta + \frac{1-\epsilon}{2} \sinh \theta \right),
 %\\
%A    \sinh \left( \frac{\nu}{2 \gamma} \mathcal{R}  \right) +B   \cosh \left( \frac{\nu}{2 \gamma} \mathcal{R}  \right)
%= \rho \left( \frac{1+\epsilon}{2} \sinh \theta + \frac{1-\epsilon}{2} \cosh \theta \right).
%\end{split}
%\label{Q10}
%\end{align}
%Here $\epsilon = \pm 1$ is  defined  as in \eqref{D61}.
%Solving \eqref{Q10} for $A$ and $B$ we obtain
%\begin{align}
%\begin{split}
%A= \rho \left[
%\frac{1+\epsilon}{2} \cosh \left( \theta- \frac{\nu}{2 \gamma} \mathcal{R} \right)+
%\frac{1- \epsilon}{2} \sinh \left( \theta- \frac{\nu}{2 \gamma} \mathcal{R} \right)
%\right],
%\\
%B= \rho \left[
%\frac{1+\epsilon}{2} \sinh \left( \theta- \frac{\nu}{2 \gamma} \mathcal{R} \right)+
%\frac{1- \epsilon}{2} \cosh \left( \theta- \frac{\nu}{2 \gamma} \mathcal{R} \right)
%\right].
%\end{split}
%\label{Q11}
%\end{align}
we note that 
since $C$, $D$, and $\rho$ are time-independent, 
 the argument of the exponentials in \eqref{W2} is a conserved quantity.
Therefore
\be
H=-2\gamma\theta+\nu\mathcal{R}
\label{Hn}
\ee
provides us with the second integral of motion for the system (\ref{St-1}).  We will employ $H$ as its Hamilton function.

Proceeding to the construction of the 
  momentum $P_\theta= {\mathcal P}$ canonically conjugate to $\theta$, we note that
  the constancy of the difference  $\nu\mathcal{R} -2\gamma \theta$
  together with the equation \eqref{S4} yield
 ${\dot \theta}=\nu Z$.
 Comparing this to the canonical equation \eqref{the},
 we conclude that the variable ${\mathcal P}$ should be introduced so as to satisfy the rule \eqref{D27}
 --- as in section \ref{om_posi} where we considered the case $\omega^2 >0$.

When  $\omega^2$ was considered positive,
the rule \eqref{D27}, the decomposition \eqref{D15},
 and  the identity $X^2+Y^2+Z^2 ={\mathcal R}^2$ were
  the only relations necessary to derive the representations \eqref{R100},  \eqref{R120},  and \eqref{R30}
  for the momentum ${\mathcal P}$.
 Also used was the fact that $x$ and $Y$ were ${\mathcal P}$-independent.
When $\omega^2$ is taken to be negative, equations \eqref{W4} give
\[
Y= 
 \frac{\nu   }{\beta_2}
 \frac{ s_2 e^{-\theta} - s_1 e^\theta}{2} 
\rho
   -\frac{4 \gamma \beta_3}{\nu^2};
\]
that is, $Y$ remains to be ${\mathcal P}$-independent.
As for the $x$, we define it by
  \[
x =              \frac{2}{\beta_2+4\beta_4}+\frac{s_1 e^\theta  +  s_2 e^{-\theta}}{2}   \rho
\]
instead of \eqref{X0}. This
preserves the validity of the decomposition \eqref{D15} --- and therefore, of
 the representations \eqref{R100}, \eqref{R120}, and \eqref{R30} for the momentum $P_\theta$.
As in the case $\omega^2>0$,  equations  \eqref{R100}, \eqref{R120} and \eqref{R30}  pertain to $|\sigma|$ smaller, greater and equal to 1, respectively.

This completes the Hamiltonian formulation of the $(\omega^2<0)$-straight 
gradient dimer  in the part of the phase space with $\rho >0$.
As in the case of the $(\omega^2>0)$-subfamily, the canonical equations are given by \eqref{Hams}.  
% As in the case $\omega^2>0$, the dimer \eqref{D10} has two independent conserved quantities,
% $\rho$ and $H$, and hence represents a completely integrable Hamiltonian system.

\subsection{$\rho=0$:
one-dimensional motion}

It remains to  consider the invariant manifold  $\rho=0$. The manifold is described by equations \eqref{D92} with
 $C=0$ or $D=0$ (supplemented by  \eqref{DX3} for the vertical coordinate).
 Assume, for definiteness, that $D=0$. Then $C$ and $\mathcal R$ define a pair of curvilinear coordinates on the 
 manifold --- which is, therefore, a two-dimensional surface.
 We will show that  the coordinate curve
 corresponding to each particular value of  $C$, 
  is a trajectory of  a Hamiltonian system with one degree of freedom.

Letting $D=0$, equations \eqref{D92} become
\be
X= C \exp \left( \frac{\nu}{2 \gamma} \mathcal R \right) + \frac{2- 2 \beta_3 \mathcal R}{\beta_2+4 \beta_4},
\quad
Y= -\frac{\nu}{\beta_2} C  \exp \left( \frac{\nu}{2 \gamma} \mathcal R \right) -
\frac{ 4 \gamma \beta_3}{\nu^2}.
\label{Q8}
\ee
The second equation in \eqref{Q8} implies that 
\[
H= \left( Y+ \frac{ 4 \gamma \beta_3}{\nu^2} \right) \exp \left( - \frac{\nu}{2 \gamma} \mathcal R \right)
\]
is a first integral of the system. We choose $\mathcal R$ as the canonical coordinate
and appoint $H$ as the Hamiltonian: $H=H(\mathcal R, \mathcal P)$.  Here $\mathcal P$ is the momentum
canonically conjugate to $\mathcal R$ (still to be introduced).

The momentum should be defined so that the canonical equation ${\dot {\mathcal R}}= \partial H/ \partial \mathcal P$
reproduce
the equation \eqref{S4}. The two equations coincide if
\be
Z= \frac{1}{2\gamma} \frac{\partial Y}{\partial \mathcal P} \exp \left( -\frac{\nu}{2 \gamma} \mathcal R \right).   \label{Q10}
\ee
Eliminating $C$ between  two equations in \eqref{Q8} we can express $X$ as
\be
X= - \frac{\beta_2}{\nu} Y +  x,      \label{Q9}
\ee
where
\be
x= \frac{2}{\beta_2+ 4\beta_4} - \frac{4 \gamma \beta_2 \beta_3}{\nu^3} -\sigma \mathcal R
\label{Qx0}
\ee
is independent of $Y$.
Assume, for definiteness, $\beta_2 \beta_4 < 0$. (The case $\beta_2 \beta_4> 0$ can be dealt with in a similar way.)
Substituting \eqref{Q10} and \eqref{Q9} in the identity $X^2+Y^2+Z^2= {\mathcal R}^2$,
 we obtain 
\be
\left( Y+ \frac{\nu}{4 \beta_4} x \right)^2+ \eta^{-2} \left( \frac{\partial Y}{\partial \mathcal P} \right)^2=r^2, \label{Q11}
\ee
where
\be
\eta^2= -\frac{16 \gamma^2 \beta_2 \beta_4}{\nu^2}   \exp \left(   \frac{\nu}{\gamma} \mathcal R \right),
\quad
r^2=   \left(  \frac{\nu}{4 \beta_4} \right)^2 x^2
 -\frac{\nu^2}{4 \beta_2 \beta_4} {\mathcal R}^2.
 \label{Qr2}
\ee
Since neither of $x$, $\eta$, or $r$ depends on $Y$, \eqref{Q11} is a separable differential equation for $Y(\mathcal P)$.
A particular solution is
$Y=  r \sin (\eta \mathcal P)-  \nu (4 \beta_4)^{-1} x$.
This relation defines $\mathcal P$: 
\[
\mathcal P=  \frac{1}{\eta} \arcsin \left[  \frac{1}{r}  \left( Y +  \frac{\nu}{4 \beta_4} x\right)  \right].
\]

Lastly, we write the Hamilton function in terms of the canonical variables:
\[
H(\mathcal R, \mathcal P)= \left[
r \sin \left( \eta \mathcal P \right)  -
\frac{\nu}{4 \beta_4} x + \frac{ 4 \gamma \beta_3}{\nu^2} \right] \exp \left( - \frac{\nu}{2 \gamma} \mathcal R \right),
\]
where the coefficients $x (\mathcal R)$, $\eta(\mathcal R)$, and
$r(\mathcal R)$  are as in   \eqref{Qx0} and
\eqref{Qr2}. The two-dimensional manifold $\rho=0$ consists of trajectories of the Hamiltonian system
${\dot {\mathcal R}}= \partial H/ \partial \mathcal P$, ${\dot {\mathcal P}}=-\partial H/ \partial \mathcal R$.
Individual trajectories are only different in the value of  $H$.

\section{Singular straight-gradient dimer: $\omega^2 = 0$}
\label{sing}

Finally we discuss the class  of dimers with $\omega^2=0$. [We remind  that $\omega^2=\beta_2(\beta_2+ 4\beta_4)$.]
The straight-gradient dimer with $\beta_2=0$ is gauge-equivalent to a cross-gradient system with $u$ and $v$ as canonical variables
(section \ref{sg}). Therefore it remains to consider the  case $\beta_2+ 4 \beta_4=0$ with $\beta_2 \neq 0$ only.

We start with  uncovering the Hamiltonian structure of the singular dimer with an 
arbitrary coefficient $\beta_3$
(subsection \ref{gen_sing}).  In the special case where $\beta_3=0$, the dimer 
admits an alternative, coexisting, Hamiltonian formulation.
This is considered in subsection \ref{spec_sing}.

\subsection{General singular dimer: arbitrary value of $\beta_3$} 
\label{gen_sing}

The general solution of  equations \eqref{D11} with $\omega=0$ has the form
  \begin{subequations}
\begin{eqnarray}
X  =  
- \frac{\beta_2}{2\gamma} Y_0 \mathcal{R} 
+ \frac{\beta_2}{4\gamma^2} {\mathcal R}^2 
-\frac{\beta_2 \beta_3}{12 \gamma^2} {\mathcal R}^3 
+ X_0,  \label{ZZ1}  \\
Y  =   -\frac{1}{\gamma} \mathcal R +
\frac{\beta_3}{2\gamma} {\mathcal R}^2
+ Y_0,  \label{ZZ2}
\end{eqnarray}
\label{ZZ00}
\end{subequations}
where $X_0$ and $Y_0$ are constants of integration. 
We define our first canonical variable by 
\be
y= Y+ \frac{1}{\gamma} \mathcal{R}
- \frac{\beta_3}{2\gamma} {\mathcal R}^2.
\label{V1}
\ee
According to \eqref{ZZ2}, $y$ is conserved: $y=Y_0$. 
As in the nonsingular situation, the advantage of using the first 
integral as a canonical coordinate is that the associated momentum $P_y$ drops out of the dynamics.

Appointing $\mathcal{R}$ as the second canonical coordinate, we need to
determine the expression for 
the momentum $P_\mathcal{R}$ canonically conjugate to $\mathcal R$.
To simplify the notation, we denote it ${\mathcal P}$.
Note that $X$ is not an independent variable here; instead, $X=X(y, \mathcal{R}, {\mathcal P})$.
Also note  the expression for the  $Y$ component of the Stokes vector:
\be
Y=y
-\frac{1}{\gamma} \mathcal R
+  \frac{\beta_3}{2\gamma} {\mathcal R}^2.
\label{V2}
\ee

%The  constant  $X_0$ will be employed as the Hamiltonian. More precisely, we let
We define the Hamiltonian by
\be
H= X+ \frac{\beta_2}{2\gamma} y \mathcal{R} -\frac{\beta_2}{4\gamma^2} {\mathcal R}^2 + \frac{\beta_2 \beta_3}{12 \gamma^2} {\mathcal R}^3.
\ee
The fact that $H$ is a conserved quantity follows from equation \eqref{ZZ1}: $H=X_0$.
To identify the momentum $P_{\mathcal{R}}={\mathcal P}$, we compare the Hamilton equation
\[
{\dot  {\mathcal{R}}}= \frac{\partial H}{\partial {\mathcal P}}
\]
to equation \eqref{S4}.  Since $\partial H/ \partial {\mathcal P}= \partial X/ \partial {\mathcal P}$,
this comparison yields
\be
Z= \frac{1}{2\gamma} \frac{\partial X}{\partial {\mathcal P}}.
\label{V4}
\ee
Substituting \eqref{V2} and \eqref{V4} in the identity $X^2+Y^2+Z^2={\mathcal R}^2$, we
obtain an ordinary differential equation for $X({\mathcal P})$:
\be
X^2+ \frac{1}{4 \gamma^2} \left( \frac{\partial X}{\partial {\mathcal P}}  \right)^2= r^2,
\label{V5}
\ee
where we have introduced 
\[
r (y, \mathcal{R})= \sqrt{ {\mathcal R}^2 - \left( y- \frac{1}{\gamma} \mathcal R + \frac{\beta_3}{2\gamma} {\mathcal R}^2 \right)^2}.
\]
Note that $r$ is independent of ${\mathcal P}$. 

A particular solution of \eqref{V5} is 
\[
X= r \sin (2 \gamma {\mathcal P});
\]
then 
\[
Z= r \cos (2 \gamma {\mathcal P}).
\]
These equations define the momentum $P_{\mathcal R}$:
\be
P_{\mathcal R}= {\mathcal P}= \frac{1}{2\gamma}   \arctan \left( \frac{X}{Z} \right).
\label{V8}
\ee

On the other hand, the momentum $P_y$ is defined by the canonical equation
\[
{\dot P_y} = -\frac{\partial H}{\partial y}.
\]
Since $\partial H / \partial P_y=0$, the right-hand side  is independent of $P_y$
and the momentum is recovered by a simple integration: $P_y= - \int (\partial H/ \partial y) dt$.

To complete the identification of the Hamiltonian structure of the singular dimer,
we  express the Hamilton function in canonical variables:
\[
 H(y, {\mathcal R}, \mathcal P)=
\frac{\beta_2}{2\gamma} {\mathcal R} \left( y- \frac{1}{2\gamma} \mathcal R + \frac{\beta_3}{6 \gamma} {\mathcal R}^2 \right) 
+
\sin (2 \gamma {\mathcal P}) 
\sqrt{ {\mathcal R}^2 - \left( y- \frac{1}{\gamma} {\mathcal R} +\frac{\beta_3}{2\gamma} {\mathcal R}^2 \right)^2}.
\]

%The existence of two independent integrals of motion makes the singular dimer a completely integrable system.

\subsection{Special singular dimer: $\beta_3=0$}
\label{spec_sing} 

% The general solution of \eqref{D11} has a particularly simple form:
%  \begin{subequations}
% \begin{eqnarray}
% X  =  -\frac{\beta_2 \beta_3}{12 \gamma^2} {\mathcal R}^3 + \frac{\beta_2}{4\gamma^2} {\mathcal R}^2 -
% \frac{\beta_2}{2\gamma} Y_0 \mathcal{R} + X_0,  \label{Z1}  \\
% Y  =  \frac{\beta_3}{2\gamma} {\mathcal R}^2 -\frac{1}{\gamma} \mathcal R + Y_0,  \label{Z2}
% \end{eqnarray}
% \label{Z00}
%\end{subequations}
% where $X_0$ and $Y_0$ are constants of integration. 

In this subsection, we consider a special subclass  of singular dimers where $\beta_3=0$ is satisfied along with 
$\beta_2+ 4 \beta_4=0$. 

Setting $\beta_3=0$
and  using (\ref{ZZ2}) to 
eliminate  $Y_0$ from (\ref{ZZ1}),  the solution (\ref{ZZ00}) becomes
\be
X= -\frac{\beta_2}{4 \gamma^2} \mathcal{R}^2 -\frac{\beta_2}{2\gamma} Y \mathcal{R} +X_0,
\quad
Y= -\frac{1}{\gamma} \mathcal{R} +Y_0.
\label{Q1}
\ee
% The constants $X_0$ and $Y_0$ give us two integrals of motion of the system (\ref{St-1}),
% of which we  construct the Hamiltonian $H$ and a coordinate $\rho$.
This time, we choose Hamilton's function $H$ to be a multiple of $Y_0$,
\be
H= -\gamma Y -\mathcal{R},   \label{H0}
\ee
and define  a canonical coordinate $\rho$ as a quadratic combination of $X$ and $Y$:
\be
\rho = \beta_2 Y^2-4X.   \label{D71}
\ee
Using \eqref{Q1}, one readily verifies that $\rho$ is a first integral: $\rho = \beta_2 Y_0^2- 4 X_0$.

Appointing $Y$
as the second canonical coordinate, the associated momentum $P_Y={\mathcal P}$ should be introduced so as to satisfy the canonical
equation ${\dot Y}=   \partial H/ \partial {\mathcal P}$.
The conjugate equation ${\dot {\mathcal P}}= - \partial H/ \partial Y$ will then be satisfied automatically.
Comparing  ${\dot Y}=   \partial H/ \partial {\mathcal P}$  to \eqref{S2} and making use of \eqref{H0}
gives
\be
\frac{\partial \mathcal R} {\partial {\mathcal P}}= 2 Z.  \label{D70}
\ee
With the help of  equation \eqref{D70}, the identity ${\mathcal R}^2-Z^2= X^2+Y^2$ becomes
\be
{\mathcal R}^2-\frac14 \left( \frac{\partial {\mathcal R}}{\partial {\mathcal P}} \right)^2= r^2,
\label{sep}
\ee
where $r$ stands for $\sqrt{X^2+Y^2}$.
Using  \eqref{D71},  $X$ can be expressed
in terms of the independent coordinates $Y$ and $\rho$. This means that  $r$  
is a function of $Y$ and $\rho$ --- but does not depend on ${\mathcal P}$:
\[
r^2= \frac{1}{16} (\beta_2 Y^2-\rho)^2 +Y^2.
\]
Accordingly,  equation \eqref{sep} can be considered as a differential equation for ${\mathcal R}({\mathcal P})$.

A simple solution to this separable equation is
\begin{subequations}
\be
\mathcal R = r \cosh (2 {\mathcal P}).
\label{D80}
\ee
Equation \eqref{D70} gives then
\be
Z=r \sinh (2 {\mathcal P}).
\label{D801}
\ee
\label{RZ}
\end{subequations}
The relations \eqref{RZ}  define the momentum: $\mathcal P= \frac12 \mathrm{arctanh} \, (Z/ \mathcal R)$.

Thus the  singular dimer  with $\beta_3=0$ and 
$\beta_2+ 4 \beta_4=0$ is a Hamiltonian system with the Hamilton function
\[
H=  -\gamma Y -  \frac{\cosh (2 {\mathcal P})}{4}
 \sqrt{   (\beta_2 Y^2- \rho)^2 +16Y^2},
\]
canonical coordinates $\rho$ and $Y$,   canonical momentum $P_Y={\mathcal P}$ defined by \eqref{RZ}, and
the momentum $P_\rho$ recoverable from ${\dot P_\rho} =- \partial H/ \partial \rho$.
This Hamiltonian formulation coexists with the formulation derived in the previous subsection.

\section{Nonlinearity-induced $\mathcal{PT}$-symmetry restoration}
\label{ST}

That some conservative systems have  all  their trajectories confined to a finite 
part of the phase space, is  a common knowledge. 
The harmonic oscillator provides a textbook example of this behaviour.

 Systems with balanced gain and loss  
 may have a similar property. 
Assume, for instance,  that the amplitudes $u$ and $v$ in \eqref{D8} and \eqref{D10}
 are small. Then 
 these $\mathcal{PT}$-symmetric dimers reduce to a two-site linear  Schr\"odinger equation:
\[
i {\dot u} + v = i \gamma u,
\quad
i {\dot v} + u = -i \gamma v.
\]
When $\gamma$ is small, all solutions to this system are bounded
but as $\gamma$ exceeds the critical value of $\gamma_c=1$,
generic initial conditions lead to solutions that grow  exponentially (with 
the growth rate $\lambda= \sqrt{\gamma^2-1}>0$). It is common to say that the $\mathcal{PT}$ symmetry is spontaneously
broken in the domain $\gamma \ge \gamma_c$ (where solutions blow up) and unbroken in the region $\gamma < \gamma_c$
(where all trajectories are confined).
The system is said to undergo the $\mathcal{PT}$-symmetry breaking transition as $\gamma$ is raised through $\gamma_c$
\cite{Guo,NatPhys,Ramezani}.

Adding  nonlinear terms may bring about a variety of effects.
Thus, the on-site nonlinearity of the standard dimer \eqref{A3}  promotes the blow-up. 
In this system, large enough initial conditions  trigger exponential growth regardless of the value of $\gamma$ \cite{pel1,flach}; furthermore,
when  $\gamma \ge 1$, {\it all} generic initial conditions  blow up \cite{flach,susanto}.
In contrast, the nonlinear coupling  of the cross-gradient dimer \eqref{A1}   softens the symmetry-breaking transition. 
In this case stable bounded solutions persist for arbitrarily large values of the gain-loss coefficient \cite{BG}.
(A similar effect is exhibited by couples of  damped-antidamped anharmonic oscillators \cite{K_IJTP}
and
solitons in a defocusing nonlinear trap with symmetrically distributed 
gain  and loss
\cite{Mal_Kart};
hence the nonlinear softening is a general phenomenon not limited to dimers.)

In what follows, we show 
that there are 
several classes of cross-gradient and straight-gradient
$\mathcal{PT}$-symmetric dimers that confine all their trajectories --- {\it regardless of the value of the gain-loss parameter $\gamma$}. 
In  these cases the nonlinearity not just softens the symmetry-breaking transition but suppresses it completely.
The 
$\mathcal{PT}$-symmetry becomes spontaneously restored.

The spontaneous symmetry restoration employs the same mechanism as
 the transition softening --- just in a more efficient way.
 The
exponential growth of small initial conditions is curbed by
the nonlinear coupling which diverts increasingly large amounts of energy from the
gaining to the losing site.
As a result, the  blow-up is arrested and
all trajectories remain  trapped
in a finite part of the phase space.
(Previously, a similar blow-up suppression was observed in a damped-driven dimer without the $\mathcal{PT}$ symmetry, namely, in the actively coupled  waveguide
pair \cite{flach,AC}.)

\subsection{Cross-gradient dimer}
\label{ST1}

We start with the family of the cross-gradient dimers \eqref{D8}.
Each member of the family
 has two independent constants of motion, $X$ and $\frak H$,
where
the Hamiltonian \eqref{D7}  has the following expression in terms of the Stokes variables:
\be
\frak H= \alpha_1 {\mathcal R}^2 + \left( \frac{\alpha_2}{4} + \alpha_4 \right) X^2 +\alpha_3 XR + \frac{\alpha_2}{4}Y^2 -\mathcal R- \gamma Y.
\label{E11}
\ee

Assume, first, that  $\alpha_2 \le 0$ while $\alpha_1+\frac14 \alpha_2 \ne 0$.  Noting that
$|Y| \le \mathcal R$, we obtain a lower bound for $\frak H$:
\be
\frak H \ge
\left(\alpha_1 - \frac{|\alpha_2|}{4} \right) \left[ \mathcal R +
\frac{\alpha_3X-\gamma-1}{2(\alpha_1+\frac14 \alpha_2)}
\right]^2 + \left( \alpha_4 + \frac{\alpha_2}{4} \right) X^2 -\frac{(\alpha_3 X-\gamma-1)^2}{\alpha_2+ 4 \alpha_1}.
\label{E12}
\ee
If $\alpha_1$ is greater than
$\frac14 |\alpha_2|$, then, keeping in mind that $X$ is a first integral,
this inequality implies that $\mathcal R$ is bounded from above by two constants of motion. That is, 
there exists an $\mathcal{R}_0$ such that $\mathcal{R}(t)  \le \mathcal{R}_0$ for all $t \ge 0$
---  the trajectory is trapped in a finite part of the phase space.

To find the Hamiltonian's lower bound in the situation where $\alpha_2>0$, we first write  Eq.\eqref{E11} in the form
\[
\frak H = \alpha_1 \left( \mathcal R + \frac{\alpha_3 X-1}{2 \alpha_1} \right)^2+ \frac{\alpha_2}{4}  \left( Y- \frac{2 \gamma}{\alpha_2}  \right)^2
+\left( \frac{\alpha_2}{4} + \alpha_4 \right)X^2 - \frac{(\alpha_3X-1)^2}{4 \alpha_1} - \frac{\gamma^2}{\alpha_2}.
\]
This representation is valid if neither $\alpha_1$ nor $\alpha_2$ is zero.
Assuming $\alpha_2 > 0$,  this gives a lower bound different from \eqref{E12}:
\[
\frak H \ge \alpha_1 \left( \mathcal R + \frac{\alpha_3 X-1}{2 \alpha_1} \right)^2 +   \alpha_4 X^2 -       \frac{(\alpha_3X-1)^2}{4 \alpha_1} - \frac{\gamma^2}{\alpha_2}       .
\]
If, in addition,  $\alpha_1>0$, this  inequality implies that there is $\mathcal{R}_0$ such that $\mathcal{R}(t) \le \mathcal{R}_0$ for all $t \ge 0$.

If $\alpha_2 < 0$ and $\alpha_1<0$, or if $\alpha_2 \geq 0$ and $\alpha_1< -\frac14 \alpha_2$,
we can establish the boundedness of $\mathcal{R}(t)$ by considering the lower bound for the integral $- \frak H$ instead of $\frak H$.

In summary,  trajectories  of the cross-gradient dimer are confined if either (a) $\alpha_1$ and $\alpha_2$ are both nonzero and have the same sign;
 (b) $\alpha_1$ and $\alpha_2$  are both nonzero and  of the  opposite sign, with $|\alpha_2| < 4 |\alpha_1|$;
 (c) $\alpha_1 \ne 0$ while  $\alpha_2=0$.

\subsection{Straight-gradient dimer}
\label{ST2}

Similar analysis can be carried out for the straight-gradient dimer \eqref{D10}.
Let, first,  $\omega^2>0$ and assume that $\mathcal{R} \to \infty$
as $t$ tends to infinity or
approaches some finite value $t_0$.
Equation \eqref{DX}  gives
\[
X= - \sigma \mathcal{R} + O (\mathcal{R}^0 ) \quad \mbox{as} \ \mathcal R \to \infty.
\]
This is only consistent with the inequality $|X| \le \mathcal R$ if $|\sigma|   \leq 1$.
Consequently, if $|\sigma|  > 1$, all trajectories of the straight-gradient dimer have to be
confined: $\mathcal R (t) \leq \mathcal{R}_0$ with some finite $\mathcal{R}_0$.

This conclusion is illustrated by the left column of Fig \ref{F1} which shows the surface \eqref{rho} with
$|\sigma |<1$,
$|\sigma| >1$,
and $|\sigma|=1$.
The surface is only seen to be compact in the middle panel, where $|\sigma| >1$.

Turning to the  $\omega^2<0$ subfamily and
assuming $\mathcal{R} \to \infty$, Eq.\eqref{D91} indicates that
the coordinate $X$ will grow exponentially in $\mathcal R$ if $C$ is nonzero.
This is clearly  inconsistent with $|X| \le \mathcal{R}$.
The only trajectories in \eqref{D91} that are consistent with this inequality, are those with $C=0$;
here the growth becomes linear in $\mathcal R$ as $\mathcal R \to \infty$.
However if $|\sigma|  >1$,
the slope of the asymptote of $X=X(\mathcal R)$  will be greater than 1. This is, again, incompatible with
$|X| \le \mathcal{R}$.
Therefore in the case $|\sigma| >1$
 all trajectories  have to be
bounded: $\mathcal R (t) \leq \mathcal{R}_0$.

The surfaces   \eqref{rho} with $\omega^2<0$ are plotted in Fig \ref{F1}, right column.
As in the left column, here we illustrated  $|\sigma| <1$,
$|\sigma| >1$,
and $|\sigma|=1$.
Only the surface with $|\sigma| >1$ (middle panel) is compact.

\begin{figure}
 \includegraphics[height=61mm, width =0.42\linewidth]{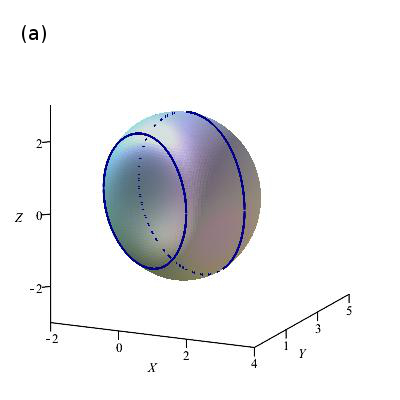}
  \includegraphics[height=61mm, width =0.42\linewidth]{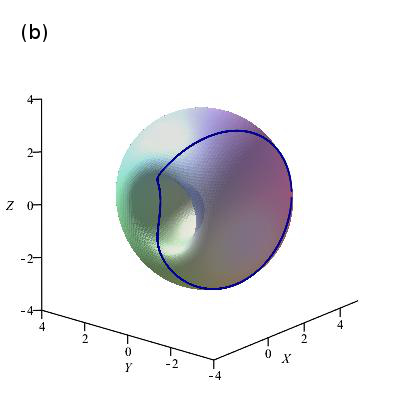}
   \caption{\label{F2}
    (a)  The surface \eqref{Z3}  with $\beta_2 \beta_3 \ne 0$.
     In this plot,
     $\gamma=1$, $\beta_2=4$, $\beta_3=1$, $\beta_4=-1$, and $X_0=3$.
     Also shown are
     two trajectories of the system \eqref{St-1}.
     (b) The surface \eqref{ZZ1} with $\beta_3=0$.
     Here, $\gamma=1$, $\beta_2=4$, $X_0=3$ and $Y_0=2$.
     This surface carries only one trajectory of the system \eqref{St-1} (depicted).
      }
\end{figure}

It remains to consider the situation
$\omega^2 =0$ which consists of two cases: (a) $\beta_2=0$ and
(b) $\beta_2+4\beta_4=0$.
In  case (a), equations \eqref{D11} yield
\[
X=X_0,
\quad
Y= \frac{ (\beta_2+4 \beta_4)X_0-2}{2 \gamma} \mathcal R
+ \frac{\beta_3}{2 \gamma} {\mathcal R}^2.
\]
As $\mathcal R \to \infty$, the expression for $Y$ is only consistent with $|Y| \leq \mathcal R$ if $\beta_3 =0$.
Therefore if $\beta_3 \ne 0$, the trajectories have to be confined.
This is in agreement with the fact that the straight-gradient dimer with $\beta_2 =0$ is gauge-equivalent to
a cross-gradient dimer with $\alpha_1 = \frac12 \beta_3$ and $\alpha_2=0$,
while
the cross-gradient dimer with $\alpha_2=0$ and $\alpha_1 \ne 0$ has been shown to suppress the blowup (see the 
previous subsection).

Finally,   we let $\beta_2+4 \beta_4=0$. To avoid the duplication of results of the previous paragraph, we also require $\beta_2 \ne 0$.
Assuming
$\mathcal R \to \infty$, 
 the exact solution (\ref{ZZ00}) indicates that 
the  $X$ 
component  would have to grow  cubically or quadratically 
 in $\mathcal R$ if $ \beta_3 \ne 0$ or
$\beta_3=0$, respectively.
However, neither cubic nor quadratic growth is consistent with the inequality $|X| \leq \mathcal R$.
Hence in the case where $\beta_2+ 4 \beta_4=0$ but  $\beta_2 \ne 0$,
all trajectories have to be bounded: $\mathcal R (t) \leq \mathcal{R}_0$.

To illustrate this conclusion geometrically, we  depict the solution 
surface for the case $\beta_2 + 4 \beta_4 = 0$. Eliminating $Y_0$ between \eqref{ZZ1} and \eqref{ZZ2} gives
\be
X-\frac{\beta_2 \beta_3}{6\gamma^2} {\mathcal R}^3 +
\frac{\beta_2}{4 \gamma^2} {\mathcal R}^2 + \frac{\beta_2}{2 \gamma} \mathcal R Y = X_0.
\label{Z3}
\ee
For each $X_0$, equation \eqref{Z3} describes a surface in the $(X,Y,Z)$ space
which hosts a one-parameter family of trajectories.
If $\beta_3 \ne 0$ and $\beta_2 \ne 0$, the surface is compact [Fig \ref{F2} (a)] so all trajectories are  confined to a finite part of the space.

When $\beta_3=0$, the surface \eqref{Z3} with a sufficiently large $\gamma$ ($\gamma> \frac12$)
 is noncompact; yet the self-trapping
of trajectories can be illustrated in this case as well.
To this end,
we note that  the equation \eqref{ZZ1}  also defines a surface in the $(X,Y,Z)$ space.
For the given $X_0$ and $Y_0$, this surface hosts a single trajectory of the dimer.
When $\beta_3=0$ (while $\beta_2 \ne 0$),
the surface \eqref{ZZ1} is compact --- for any $\gamma$ and any choice of $X_0$ and $Y_0$.
This is illustrated in Fig \ref{F2}(b).

To summarise, trajectories  of the straight-gradient dimer are confined if  either (a) 
$\beta_2 (\beta_2 + 4 \beta_3) \neq 0$ and $2 |\beta_3| > |\beta_2 + 4 \beta_4|$;
(b) $\beta_2=0$ but $\beta_3 \ne 0$; or (c) $\beta_2+ 4 \beta_4=0$ but $\beta_2 \ne 0$.

%The standard dimer
%\eqref{A3} has all initial conditions blowing up in the region $\gamma \ge 1$ \cite{flach}.
%Therefore, its point of the nonlinear $\mathcal{PT}$-symmetry breaking transition is the same as the linear transition point:
%$\gamma_c(P)=\gamma_c(0)$.
%In contrast, the critical point of the cross-gradient dimer \eqref{A1} does depend on $P$:
%$\gamma_c(P)=1+P$ \cite{BG}.
%In this case, the nonlinearity softens the symmetry-breaking transition.
%(A similar softening is exhibited by couples of  damped-antidamped anharmonic oscillators \cite{K_IJTP}
%and
%solitons in a defocusing nonlinear trap with symmetrically distributed
%gain  and loss
%\cite{Mal_Kart}.)

%\section{Remarks on possible applications and Conclusions}

\section{A note on applications}
\label{appli}

The standard dimer \eqref{A3}  describes the coupling of a Kerr optical waveguide with loss to a twin waveguide characterised by the optical gain 
of  equal rate  \cite{Maier,CSP,PT_Opt,Guo,NatPhys,Ramezani}. 
In the boson condensation context, the same pair of
equations governs condensates  in two identical potential wells \cite{BEC}, with one well losing 
and the other one being fed with atoms \cite{Graefe,Heiss}.
The cross-gradient dimer \eqref{A1} 
%and some other representatives of the families \eqref{D8} and \eqref{D10}
models the same potential-well geometry of the condensate, the difference from the standard dimer
being that this time $u$ and $v$ are the amplitudes of the symmetric and antisymmetric state \cite{Ostrovskaya}
rather than the amplitudes of the condensate in the left and right well.

The aim of this section is to emphasise that other 
nonlinear models in \eqref{D8} and \eqref{D10} are not physically irrelevant either.
In particular, 
these dimers furnish amplitude equations for
couples of  oscillators with physically realistic nonlinearities. 
We exemplify this correspondence by simple systems with the gain-loss balance of.two 
different types.

\subsection{Two pendula  with periodic coupling}

The first  system consists of two pendula with
 a periodically varied coupling:
\be
\frac{d^2 x}{d \tau^2}  + \sin x + \kappa(\tau) y= 0,         \quad
 \frac{d^2 y}{d \tau^2}+ \sin y + \kappa(\tau)  x = 0.
\label{par}
\ee
The coupling is assumed to be weak and varied at the frequency close to the double natural
frequency of each pendulum:
\[
\kappa = 2 \epsilon^2 \cos( 2 \omega \tau), \quad \omega = 1 - \Omega \epsilon^2.
\]
Here $\epsilon^2$ is a small parameter that sets the scale of the amplitude of the coupling
 modulation,
while the coefficient $\Omega=O(1)$   measures the detuning 
of the driving half-frequency from the  frequency of the linear oscillations.

This type of parametric driving can be easily realised experimentally. For example, 
the pendula can be hung  from a common horizontal rope,
with a periodically varied rope tension.

Assuming that the pendula are performing small-amplitude librations, we expand $x$ and $y$ 
 in powers of $\epsilon$:
\[
x = \epsilon x_1 + \epsilon^3 x_3 +..., \quad
y = \epsilon y_1 + \epsilon^3 y_3 + ...
% \label{expansion}
\]
The coefficients of the expansion are
allowed to depend on a hierarchy of  time  scales $T_n=\epsilon^n \omega \tau$, $n=0,2, ...$. 
The times $T_n$ become independent as $\epsilon \to 0$; in this limit, $d/d\tau= \omega(D_0+ \epsilon^2 D_2 + ...)$, where 
$D_n = \partial / \partial T_n$.  We also develop $\sin x$ and $\sin y$ in powers of their arguments.

Substituting these expansions in \eqref{par} 
and equating coefficients of like powers of $\epsilon$, the order $\epsilon^1$ yields
\[
x_1= Ae^{i T_0} + A^* e^{-i T_0}, \quad y_1= iBe^{i T_0} - iB^* e^{-iT_0},
\]
where $A$ and $B$ are functions of $T_2, T_4, ...$ ---  but not of $T_0$. 
At the order $\epsilon^3$ we obtain a pair of equations
\begin{align*}
(D_0^2+1)x_3 = 2\Omega D_0^2 x_1 -2 D_0D_2 x_1 - 2  \cos (2T_0) y_1+ \frac16 x_1^3,  \\
(D_0^2+1)y_3 = 2\Omega D_0^2 y_1 -2 D_0D_2 y_1 - 2  \cos (2T_0) x_1 + \frac16 y_1^3.
\end{align*}
Substituting for $x_1$ and $y_1$, and setting the secular term equal to zero we arrive at
\begin{align} \begin{split} 
2iD_2 A + 2 \Omega A -i B^*  - \frac12 |A|^2A=0,    \\
2i D_2 B + 2 \Omega B -i A^*  - \frac12 |B|^2 B=0.  \label{2N}
\end{split} 
\end{align}

Assuming, for definiteness, $\Omega >0$ 
and letting
\[
A= 2\sqrt{\Omega} (u+v), \quad B= 2 \sqrt{\Omega} (u^*- v^*), 
\]
equations \eqref{2N}  become
\begin{align} \begin{split} 
i {\dot u} +v - i \gamma u= u^2v^* + (2|u|^2+ |v|^2) v,    \\
i {\dot v}  + u + i \gamma v =   v^2u^* + (2|v|^2+|u|^2) u,  \label{3N}
\end{split} 
\end{align}
where we have introduced  $ \gamma= (2\Omega)^{-1}$ and 
the overdot indicates differentiation with respect to 
$t= \Omega  T_2$.
The system \eqref{3N} is nothing but the cross-gradient 
$\mathcal{PT}$-symmetric dimer \eqref{D8}
with $\alpha_2=\alpha_3=0$ and $\alpha_1=\alpha_4=\frac12$.

\subsection{Damped-antidamped oscillator couples}

Another, unrelated, interpretation of equations \eqref{D8} and \eqref{D10} is that of
 amplitude equations for a damped oscillator
coupled to an oscillator with negative damping.
(This time the coupling coefficient is assumed to be constant.)
Following the multiple-scale procedure of the previous subsection,
one can readily check that
 the cross-gradient dimer \eqref{D8} governs the oscillation amplitudes of the
damped-antidamped pair
\begin{align} \begin{split}
x_{\tau \tau}  + \eta x_\tau +x + \kappa y = c_0(x^2+ 3y^2)x  + (c_1 x^2+c_2 y^2)y,         \\
y_{\tau \tau}   - \eta y_\tau +y + \kappa x  = c_0 (y^2+ 3x^2) y  + (c_1 y^2+c_2 x^2)x. \label{D800}
\end{split}
\end{align}
Here $\kappa=  \epsilon^2$ and
$\eta = \gamma \epsilon^2$  are the coupling and the
gain-loss coefficient (assumed small),
while the nonlinearity coefficients $c_0$, $c_1$ and $c_2$ can be chosen arbitrarily.
The dimer parameters in  \eqref{D8} are expressible through the coefficients in \eqref{D800}:
$\alpha_1= \frac32 c_2$, $\alpha_2=c_1-3c_2$,
$\alpha_3= 3 c_0$, and $\alpha_4= \frac12 c_1$.

On the other hand,  the straight-gradient dimer \eqref{D10} serves as the amplitude system for
the oscillator couple
\begin{align*}
x_{\tau \tau}  + \eta x_\tau +x + \kappa y = (c_1 x^2+ c_2y^2)x + c_0 (3x^2+y^2) y,         \\
y_{\tau \tau}   - \eta y_\tau +y + \kappa x =   (c_1 y^2+ c_2 x^2)y  +c_0 (3y^2+x^2)x. 
\end{align*}
This time, the relation between the coefficients of the oscillators and the dimer parameters
is as follows:
$\beta_1=\frac32 c_1$, $\beta_2= c_2- 3c_1$, $\beta_3= 3c_0$, and $\beta_4=\frac12 c_2$.

The damped-antidamped oscillator model has been employed to interpret experiments in systems
as diverse as a tied pair of magnetically kicked pendula \cite{pendula}, two connected optical whispering galleries \cite{Benfreda,MG_Nature},
or a tandem of inductively coupled active \textit{LRC} circuits --- one with amplification and the other one attenuated at an equal rate \cite{Schindler}.

\section{Conclusions}
\label{conclusions}

The main results of this study can be summarised as follows.

(1)
We have introduced a four-parameter
($\alpha_1, \alpha_2, \alpha_3, \alpha_4$)
family of Hamiltonian $\mathcal{PT}$-symmetric dimers with a cross-gradient canonical structure,
equations \eqref{D8}. The entire family was shown to have an additional 
integral of motion, independent of the Hamiltonian; 
hence each member of the family is 
 Liouville integrable.
%We have furnished an analytic description of all trajectories in these systems.
All trajectories in these systems were described analytically.

(2)
We have considered a $\mathcal{PT}$-symmetric extension of 
a four-parameter ($\beta_1, \beta_2, \beta_3, \beta_4$)
family of  
conservative dimers with the straight-gradient Hamiltonian structure, equations \eqref{D10}.
 Unlike for the cross-gradient dimers, the original complex $u, v$ variables
do not  constitute canonical coordinates for the straight-gradient family \eqref{D10}
--- except when the straight-gradient dimer has a coexisting cross-gradient formulation.
 The three-parameter ($\beta_1, \beta_3, \beta_4$) subfamily of
 straight-gradient dimers  with $\beta_2=0$ was shown to 
admit  such an alternative cross-gradient representation.

We have identified canonical coordinates and momenta
and uncovered the Hamiltonian structure for the entire four-parameter family of the
 straight-gradient $\mathcal{PT}$-symmetric dimers.
 By establishing that 
 each member of the family has an additional integral of motion, we
  have demonstrated that the entire family is Liouville integrable.
All trajectories of the straight-gradient $\mathcal{PT}$-symmetric dimers were described analytically.

(3)
We have proved that 
the cross-gradient dimer with parameters $\alpha_{1,2}$ satisfying
(a) $\alpha_1 \alpha_2>0$;
or (b) $\alpha_1 \alpha_2<0$ with $|\alpha_2| < 4 |\alpha_1|$;  or
(c) $\alpha_1 \ne 0$ while  $\alpha_2=0$ ---
has all trajectories bounded, irrespectively of the values of other parameters $\alpha_{3,4}$ or the gain-loss coefficient $\gamma$.

(4)
We have established that 
regardless of the value of  $\beta_1$ and the gain-loss coefficient $\gamma$,
the straight-gradient dimer with parameters satisfying  (a)
$\beta_2 (\beta_2 + 4 \beta_3) \neq 0$, $2 |\beta_3| > |\beta_2 + 4 \beta_4|$;
or (b) $\beta_2=0$, $\beta_3 \ne 0$,  with no constraints on $\beta_4$;
or (c) $\beta_2+ 4 \beta_4=0$, $\beta_2 \ne 0$, with no constraints on $\beta_3$ ---
has all trajectories bounded.

(5)
We have demonstrated that the amplitudes of libration of
 two coupled oscillators with on-site  nonlinearities, 
driven by the periodic variation of their coupling coefficient,  
satisfy a  $\mathcal{PT}$-symmetric cross-gradient dimer system. $ \Box$ \\

Thus, the principal mathematical  conclusion  is that 
 the $\mathcal{PT}$-symmetric extensions of 
{\em all} conservative nonlinear Schr\"odinger dimers  remain
 completely integrable  Hamiltonian             systems. On the other hand, the principal  physical upshot
 is that 
there are broad classes of
$\mathcal{PT}$-symmetric dimers that confine all their trajectories {\it regardless of the value of the gain-loss parameter $\gamma$}. 
The
$\mathcal{PT}$-symmetry, which is broken at the level of  the underlying linear equation, 
becomes spontaneously restored thanks to the nonlinear coupling.

The spontaneous $\mathcal{PT}$-symmetry restoration may find applications in 
integrated optics where $\mathcal{PT}$-symmetric 
nonlinear Schr\"odinger dimers describe directional waveguide couplers.
A nonlinear coupler composed
of one core with a certain amount of optical gain and another one
with an equal amount of loss switches the entire power to one waveguide
\cite{CSP}.
In the standard  $\mathcal{PT}$ dimer, this  power switching is accompanied by an
unbounded power growth in one of the arms of the
device ---
the growth  not saturable  by nonlinearity \cite{SXK,PT2,NatPhys,Ramezani}.
In contrast,  no input can trigger an uncontrollable
growth of optical modes in a dimer with the nonlinearly-restored $\mathcal{PT}$ symmetry.
As a result, 
the $\mathcal{PT}$  symmetry restoration may represent a technological advantage.

\section*{Acknowledgments}
%  We thank Boris Malomed for instructive correspondence.
This project began during one of the authors (D.E.P.'s) visit to the University of Cape Town in April 2014.
He would like to thank  the Visiting Scholars Fund of UCT for financial support.
The work of P.D. was funded through a postdoctoral fellowship from the Claude Leon Foundation.
I.B. acknowledges research funding from the NRF of South Africa (grants No 85751, 86991, and 87814).
The work of D.E.P. was supported by the Ministry of Education
and Science of Russian Federation (the base part of the state task No. 2014/133, project No. 2839).
Last but not least, it is a pleasure to thank  Vladimir Konotop, Tsampikos Kottos, Boris Malomed,  and Vyacheslav Priezzhev
for useful discussions.

%%%%%%%%%% Insert bibliography here %%%%%%%%%%%%%%

\end{document}